\documentclass[lettersize,journal]{IEEEtran}
\usepackage{amsmath,amsfonts}
\usepackage{algorithmic}
\usepackage{algorithm}
\usepackage{array,makecell,bbding,pifont}
\usepackage[caption=false,font=normalsize,labelfont=sf,textfont=sf]{subfig}
\usepackage{textcomp}
\usepackage{stfloats, threeparttable}
\usepackage{url}
\usepackage{verbatim}
\usepackage{graphicx}
\usepackage{cite}
\usepackage{url}
\usepackage{colortbl}
\usepackage{xcolor}
\usepackage{array}
\usepackage{color}
\usepackage{hyperref}  
\usepackage{upgreek}
\usepackage{enumerate}
\usepackage[figuresright]{rotating}
\usepackage{bbding}

\begin{document}

\title{Networking Architecture and Key Supporting Technologies for Human Digital Twin in Personalized Healthcare: A Comprehensive Survey}

\author{Jiayuan Chen, Changyan Yi,~\IEEEmembership{Member,~IEEE}, Samuel D. Okegbile,~\IEEEmembership{Member,~IEEE}, Jun Cai,~\IEEEmembership{Senior Member,~IEEE}, and Xuemin (Sherman) Shen,~\IEEEmembership{Fellow,~IEEE}
	
	\IEEEcompsocitemizethanks{\IEEEcompsocthanksitem J. Chen and C. Yi are with the College of Computer Science and Technology, Nanjing University of Aeronautics and Astronautics, Nanjing, Jiangsu, 211106, China. (E-mail: \{jiayuan.chen, changyan.yi\}@nuaa.edu.cn).
		\IEEEcompsocthanksitem 
		S. D. Okegbile and J. Cai are with the Network Intelligence and Innovation Laboratory (NI$^2$Lab), Department of Electrical and Computer Engineering, Concordia University, Montreal QC H3G 1M8, Canada. (Email: \{samuel.okegbile, jun.cai\}@concordia.ca).
		\IEEEcompsocthanksitem 
		X. Shen is with the Department of Electrical and Computer Engineering, University of Waterloo, Waterloo, ON N2L 3G1, Canada.
		(Email: sshen@uwaterloo.ca).\\
		\emph{(Corresponding authors: Changyan Yi and Jun Cai.)}
	}
	
}

\maketitle

\begin{abstract}
	Digital twin (DT), referring to a promising technique to digitally and accurately represent actual physical entities, has attracted explosive interests from both academia and industry. One typical advantage of DT is that it can be used to not only virtually replicate a system’s detailed operations but also analyze the current condition, predict the future behavior, and refine the control optimization. Although DT has been widely implemented in various fields, such as smart manufacturing and transportation, its conventional paradigm is limited to embody non-living entities, e.g., robots and vehicles. When adopted in human-centric systems, a novel concept, called human digital twin (HDT) has thus been proposed. Particularly, HDT allows \emph{in silico} representation of individual human body with the ability to dynamically reflect molecular status, physiological status, emotional and psychological status, as well as lifestyle evolutions. These prompt the expected application of HDT in personalized healthcare (PH), which can facilitate the remote monitoring, diagnosis, prescription, surgery and rehabilitation, and hence significantly alleviate the heavy burden on the traditional healthcare system. However, despite the large potential, HDT faces substantial research challenges in different aspects, and becomes an increasingly popular topic recently. In this survey, with a specific focus on the networking architecture and key technologies for HDT in PH applications, we first discuss the differences between HDT and the conventional DTs, followed by the universal framework and essential functions of HDT. We then analyze its design requirements and challenges in PH applications. After that, we provide an overview of the networking architecture of HDT, including data acquisition layer, data communication layer, computation layer, data management layer and data analysis and decision making layer. Besides reviewing the key technologies for implementing such networking architecture in detail, we conclude this survey by presenting future research directions of HDT.
\end{abstract}


\section{Introduction}
\subsection{Background and Motivation}
\IEEEPARstart{T}{he} inherent shortage of healthcare resources has caused ongoing challenges to the traditional healthcare system. For example, COVID-19 pandemic in 2020-2023 has resulted in surges of demands for medical facilities, such as ventilators, extracorporeal membrane oxygenation (ECMO) and testing machines, which exceeds the capacity of the traditional system, leading to tens of millions of people infected and died \cite{47}. 
Meanwhile, the cost burden on the traditional healthcare system is rapidly growing in most worldwide regions. For instance, the Centers for Medicare \& Medicaid Services predicts that U.S. healthcare spending will grow at a rate 1.1\% faster than that of the annual gross domestic product (GDP) and is expected to increase from 17.7\% of the GDP in 2018 to 19.7\% (reach to \$6.2 trillion) by 2028 \cite{6}. 
Despite healthcare expenditures being projected to increase at such a substantial rate, the traditional healthcare system produces no better (and indeed sometimes worse) outcomes. A recent analysis estimated that about one-quarter of total healthcare spending in the U.S. (between \$760 billion and \$935 billion annually) is wasteful, mainly attributed to ineffective and inefficient treatments \cite{10}.

To this end, the relentless proliferation of disruptive information technologies, such as 5G, big data, artificial intelligence (AI), have open rich opportunities for the realization of highly efficient personalized healthcare (PH) services.
PH is an innovative way that seeks to offer preventive care and targeted treatments for each individual patient through his/her unique medical records, genes and values. In other words, PH can provide a more precise treatment approach, one-size-fits-one approach, eliminating unnecessary side effects and high cost of one-size-fits-all approach widely adopted in the traditional healthcare system.
Furthermore, PH can also help medical personnel make better decisions and facilitate breakthroughs for many difficult-to-treat rare diseases given the individualized data-driven information, thereby improve the quality of life of people around the world \cite{146}.

Unsurprisingly, AI plays a prominent role in the implementation of PH. As one of the most popular research hotspots, AI has penetrated in all aspects of people's lives with strong muscle in data analysis. 
AI can fuel the progress of digitization and intelligence of all industries, particularly for healthcare, where AI emulates human cognition in the analysis of complicated medical data using complex algorithms and software.  
 To guarantee superior PH services, AI requires accurate learning and construction of many high-performance and personalized feature models for individuals, based on massive high-quality individual training datasets \cite{468}. Nevertheless, this process of acquiring individual datasets from a single person (particularly for supervised learning which needs to manually label these datasets) is significantly costly, error-prone and time-consuming. Furthermore, the existing AI-driven efforts mostly provide solutions through regression-based and classification-based approaches for real-valued and discrete-valued attributes predictions respectively, and thereby, such solutions are limited in scope to specific diseases, diagnosis or a small subset of the population \cite{49}.

 To eliminate the limitations of the current AI-driven solutions for PH, digital twin (DT), is envisioned as a promising paradigm to adopt. Specifically, DT can be seen as an exceptionally vivid testbed for replicating the condition, function and operation of each individual, and with the ability of running an unlimited number of virtual ``what if'' simulations without harming the human body. With this feature, AI integrating in DT could be trained much more efficiently with massive and diverse individual synthetic datasets generated by DT (besides those collected from physical world). Furthermore, DT can in turn utilize a myriad of high-performance and coupled AI models to assist in capturing more comprehensive and high-fidelity of the extremely complex human body system. Additionally, since the conditions of the human body system are ever-changing (e.g., with aging, behavioral and environmental changes), AI models integrating in DT can be dynamically validated and updated for precisely abstracting the real status of the human body system, and thus become much more self-adapted.

\subsection{Demystifying Human Digital Twin}
The concept of DT was proposed in 2003 and referred to the digital replica of a physical entity \cite{14}. DT is the convergence of several cutting-edge technologies, such as big data and AI, 5G/6G and Internet of things (IoT), data visualization and extended reality (XR)\footnote{XR is defined as an umbrella term that encompasses virtual reality (VR), augmented reality (AR), mixed reality (MR) \cite{467}.}, communication and computation technologies, blockchain and cybersecurity. DT is at the forefront of the Industry 4.0 revolution, and is being widely implemented in diverse areas, e.g., manufacturing \cite{16}, city transportation \cite{19}, smart construction \cite{25}, and smart wireless systems \cite{462}. These applications are mainly related to non-living physical entities. When adopted in human-centric applications and systems, a new concept, called human digital twin (HDT), has emerged, which allows an \emph{in silico} representation of any individual with the ability to dynamically reflect molecular status, physiological status, emotional and psychological status, as well as lifestyle evolutions \cite{387, 519}.

HDT is expected to play an essential role in shaping the future of healthcare systems. 
It is, therefore, unsurprising that HDT is now receiving wide attention in many healthcare industries owing to its capabilities to improve PH. To mention a few, we briefly discuss some benefits of HDT. 
\begin{itemize}
	\item  HDT can facilitate continuous monitoring of individual health status with the ability to predict viral infections and possible corresponding immune responses, thus allowing rapid and proactive interventions. 
	\item HDT can improve the efficiency of clinical trials by ensuring that clinical trials are carried out on the virtual replica of human beings, thus promoting an efficient pharmaceutical industry procedure while supporting safer drugs and accelerating vaccine development process.
	\item HDT can facilitate therapeutic plans option by recommending safe and patient-specific medical therapies based on the unique genetic profile of each individual. With this, HDT can prevent harmful side effects and improves medical outcomes while saving cost. 
	\item Through the use of current biomarkers, HDT can facilitate early identification of genomic and epigenomic events such as carcinogenesis in disease progression, thus allowing earlier detection of illness. 
	\item HDT is capable of predicting the future health condition of each individual, thereby allowing a proper activation of efficient preventive measures.
	\item HDT can reduce health inequalities through telemedicine. Patients can remotely access PH irrespective of their geographic locations.
	\item HDT can promote precisions of diagnosis. With HDT, digitized sensations of pain and anxiety are converted into a form that can be observed by the medical personnel. \end{itemize}

While HDT development is still in its infancy, many med-tech giants, such as Siemens, Philips and IBM, are currently exploring the possibility of facilitating the commercialization of HDT by relying on their massive databases and strong financial strengths. In Table \ref{table2}, we provide a glance of the current industrial progress on HDT.

\begin{table*}[htp] 
	\centering
	\caption{Industrial Progress on Human Digital Twin (HDT)} 
	\label{table2}
	\begin{tabular}{| m{5cm} <{\centering}| m{12.31cm}<{\centering} | }
		\hline
		\rowcolor{lightgray}\textbf{Company$\backslash$Project} & \textbf{Product$\backslash$Service} \\
		\hline
	Siemens Healthineers \cite{31} & 3D Digital Heart Twin -- 3D Digital Heart Twin solution is capable of making definitive and timely diagnoses while facilitating simulations of surgical procedures and treatment trials. It is one of the first full-fledged ward management DTs. \\
	\hline
	Philips \cite{32} & Heart model -- Heart model is a personalized heart DT.  \\
	\hline
	European CompBioMed project \cite{33} & The project focuses on the use and development of computational methods for biomedical applications, including the virtual human. \\
	\hline
	The SIMULIA living heart project \cite{34} &  Released by Dassault Syst\`{e}ms, the SIMULIA living heart -- a DT model – translates electrical impulses into mechanical contractions and is the first digital organ representation that possesses the same functionality as the physical organ. \\
	\hline
	IBM \cite{35} & IBM DT simulates the body’s biochemical processes through AI techniques to detect cancerous cells in any previously obtained health data. \\
	\hline
	Digitwins \cite{36} & Digitwins aims to develop a revolutionary-based approach for healthcare systems through detailed modelling processes to facilitate simulations of numerous treatments without subjecting patients to any form of harm. \\ 
	\hline 
	General Electric \cite{37} & An application called “Predix” was developed by General Electric in 2016 to create DTs for patients. “Predix” is responsible for running data analytics and monitoring. \\ 
	\hline
	Sim \& Cure \cite{38} & Sim \& Cure developed a DT that can assist surgeons when choosing appropriate endovascular implants that can optimize aneurysm repair. \\ 
	\hline
	Swedish DT Consortium (SDTC) \cite{146, 168} & The SDTC strategy for PH is based on: i) constructing multiple HDT copies of any single patient; ii) computationally treating each of these HDTs with different medications to identify the best-performing medication; iii) treating the patient with this best-performing medication. \\
	\hline
	 European ecosystem for digital twins in healthcare (EDITH) project \cite{530} & 	EDITH has built an ecosystem, where all members of the consortium aim to develop a repository of healthcare DTs, and work collaboratively to achieve the ultimate objective, i.e., the establishment of a fully-fledged HDT. \\
	\hline
		SEMARX \cite{533}& 	SEMARX has developed HDT for two use cases: i) HDT for patients, mainly for assisting neonatology ICU patients; ii) HDT for children, mainly for assuring children's safety.\\
	\hline
	\end{tabular}
\end{table*}

\subsection{Related Work}

As an underlay of HDT, DT continues to attract a myriad of attentions. In \cite{358}, Barricelli et al. provided the state-of-the-art definitions of DT including its fundamental characteristics as well as its common application areas. The survey carried out in \cite{455} 
focused on introducing  the concept of DT in wireless systems for addressing issues such as security, privacy and air interface design. In \cite{454}, the authors reviewed the framework of DT in the view of industrial IoT applications.

Following \cite{358, 454, 455}, subsequent efforts delved into more comprehensive and specific enabling technologies for DTs. For example, Rasheed et al. in \cite{153} enumerated the common challenges in DTs and surveyed the corresponding enabling technologies, such as digital platforms, cryptography, blockchain, big data, data privacy and security, data compression, 5G and IoT for real-time communication.
Alcaraz et al. in \cite{371} abstracted the architecture of DT into four layers, i.e., data dissemination and acquisition, data management and synchronization, data modelling and additional services, data visualization and accessibility, and further explored the enabling technologies that can be used to achieve each of these architectural layers.
Khan et al. \cite{164} provided a brief overview of key technologies for Industry 4.0, including IoT, big data, AI, and DT (which is the confluence of the technologies mentioned above). Additionally, this work briefly introduced tools for the construction of DT, such as tools for DT modelling and tools for data management in DT.

However, these surveys mainly focused on the implementation of DT in industries, and we call them as the conventional DT hereafter. The design requirements of HDT and the conventional DT are significantly different in many aspects, especially when considering HDT from the PH perspective. Hence, the enabling technologies of the conventional DT described in the existing surveys may not be directly adopted in HDT. Several recent studies have started to discuss enabling technologies for HDT solutions in PH applications. For instance,
Ferdousi et al. in \cite{169} presented a very-high level overview of HDT design requirements while highlighting the differences between HDT and conventional DTs. The authors briefly discussed some of the underlying technologies used in the development of a HDT. They provided a use-case scenario where a DT of a patient with a mental issue was created to monitor and predict stress levels. Similarly, El Saddik et al. \cite{170} elaborated on the architecture (consisting of data source, AI-inference engine and multimodal interaction (MMI)) and design requirements of HDT. The authors carried out an in-depth investigation of five cutting-edge technologies to support the proposed HDT in \cite{43}. These five technologies include big data technology with huge amount of data collected using IoT devices and social networks, AI algorithms to extract information, cybersecurity technology for securing the collected personal data, MMI technology for interfacing the real and virtual twin, and quality of experience (QoE) based communications for providing high performing networks. Okegbile et al. in \cite{387} provided an insight into HDT for PH services. The authors presented architectural frameworks as well as key design requirements (including model conceptualization, data representation model, scalable AI-driven analytics, model scalability and reliability, and security) of HDT and investigated the key technologies (such as connectivity, data collection, data processing, digital modelling, AI solutions for decision making and cloud-edge computing for storage and computation) with various challenges to suggest future research directions. Lin et al. \cite{475} conducted an extensive literature review on HDT, analyzing enabling technologies and establishing typical frameworks. Particularly, they focused on the sensing/perception technology and two modelling technologies of human body or organ and human behavior.

\begin{table*}[htp]
	\centering
	\caption{The Comparison of the State-Of-The-Art Survey}
	\label{table5}
	\begin{tabular}{|c|c|c|c|c|c|c|c|}
	\hline
	\rowcolor{lightgray}\textbf{Reference} & \textbf{\makecell{Targeted\\Application}} & \textbf{\makecell{Networking\\Architecture}} & \textbf{\makecell{Data\\Acquisition}} & \textbf{Communication} & \textbf{Computation} & \textbf{\makecell{Data\\Management}} & \textbf{\makecell{Data Analysis}}\\
	\hline
	\makecell{Khan et al.\\\cite{455} 2022} & \makecell{Conventional\\DT} & \XSolidBrush & \XSolidBrush & \CheckmarkBold & \CheckmarkBold  &\CheckmarkBold & \XSolidBrush \\
	\hline
	\makecell{Tao et al.\\\cite{454} 2019} & \makecell{Conventional\\DT} & \XSolidBrush & \XSolidBrush & \XSolidBrush&\XSolidBrush &  \CheckmarkBold& \XSolidBrush\\
	\hline
	\makecell{Rasheed et al.\\\cite{153} 2020}& \makecell{Conventional\\DT}& \XSolidBrush & \CheckmarkBold & \CheckmarkBold & \CheckmarkBold& \CheckmarkBold& \CheckmarkBold  \\
	\hline
		\makecell{Alcaraz et al.\\\cite{371} 2022}& \makecell{Conventional\\DT} & \CheckmarkBold& \CheckmarkBold & \CheckmarkBold & \CheckmarkBold& \CheckmarkBold & \CheckmarkBold \\
	\hline
		\makecell{Khan et al.\\\cite{164} 2022 }& \makecell{Conventional\\DT} & \XSolidBrush &\CheckmarkBold & \CheckmarkBold &  \CheckmarkBold & \CheckmarkBold&\CheckmarkBold  \\
	\hline
		\makecell{Lin et al.\\\cite{475} 2023} &HDT & \XSolidBrush& \CheckmarkBold&\XSolidBrush& \XSolidBrush& \XSolidBrush& \XSolidBrush \\
	\hline
	This work & HDT &  \CheckmarkBold& \CheckmarkBold& \CheckmarkBold& \CheckmarkBold& \CheckmarkBold& \CheckmarkBold\\
	\hline 
\end{tabular}
\end{table*}

Despite the aforementioned surveys or magazines that have discussed various aspects of HDT and the conventional DT (as summarized in Table \ref{table5}), these works neither offered the networking architecture enabling HDT in PH applications, nor delve into the key technologies for supporting the networking architecture. However, it is obvious that the practical implementation and operation of HDT heavily rely on its networking architecture. This motivates us to compose this survey that particularly discusses the networking architecture of HDT in PH applications. Through surveying the key technologies enabling the networking architecture, our survey provides critical insights and useful guidelines for the readers to better understand how to realize HDT for PH applications and discover the open issues on this topic.

\subsection{Contributions}
 We carry out a comprehensive survey on the networking architecture and key technologies for HDT in PH applications. Our survey can provide not only critical insights and useful guidelines for readers to have a better understanding of the networking architecture of HDT for PH applications, but also the key technologies for enabling such networking architecture. The contributions of this survey is a bold, forward-looking vision of HDT, and they are summarized as follows:

 \begin{itemize}
	\item We summarize an overview of HDT from the existing literature, including its differences compared to the conventional DT, a universal framework and the essence in PH applications. Through all these, readers can gain an in-depth view of this topic.
	\item Different from the existing literature, we analyze the design requirements and challenges of HDT from a novel networking perspective. This provides readers with the understanding of how HDT in PH applications can be achieved ubiquitously, timely, securely and accurately.
	\item We are the first to comprehensively investigate the end-to-end networking architecture of HDT. Specifically, we shed light on the five-layered networking architecture of HDT, including data acquisition, communication, computation, data management, and data analysis and decision making layers. Moreover, we further survey the key supporting technologies for each layer. In this way, readers with networking backgrounds can be potentially motivated to conduct more out-of-the-box research in facilitating the development of HDT.
	\item We outline several research directions to encourage future studies in this area. Our survey can serve as an initial step that precedes a holistic and insightful study of the networking architecture and Key supporting technologies for HDT in PH applications, helping researchers quickly grasp this area.
\end{itemize}

\begin{figure*}[!t]
	\centering
	\includegraphics[width=0.99\textwidth]{ 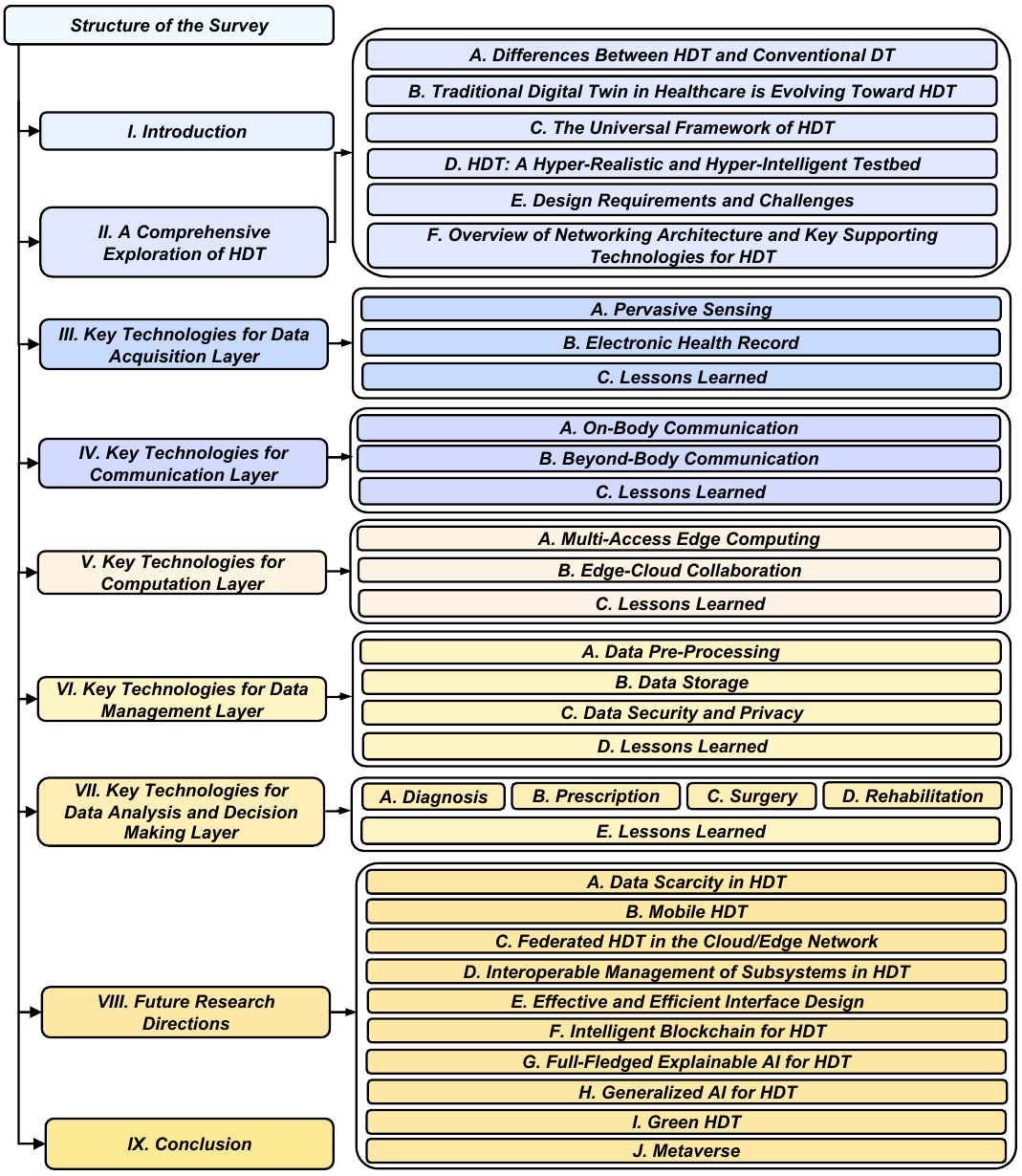} \\
	\caption{The organization of this survey.}\label{Stru}
\end{figure*}

 The structure of this survey is visualized as shown in Fig. \ref{Stru}. For convenience, Table \ref{table3} lists all common abbreviations.

\begin{table}[t]
	\caption{List of Abbreviations}
	\label{table3}
	\begin{tabular}{| m{2cm} <{\centering} | m{6cm} <{\centering}| }
		\hline
		\rowcolor{lightgray}\textbf{Abbreviation} & \textbf{Full Form} \\
		\hline
		DT & Digital twin \\
		\hline
		HDT & Human digital twin \\
		\hline
		PT &  Physical twin \\
		\hline
		VT & Virtual twin \\
		\hline
		PH & Personalized healthcare\\
		\hline
		EHR & Electronic health record \\
		\hline
		IoT & Internet of things \\
		\hline
		IoMT & Internet of medical things \\
		\hline
		ACC & Accelerometers \\
		\hline
		ECG & Electrocardiograms\\
		\hline
		EEG & Electroencephalographic\\
		\hline
		PPG & Photoplethysmography\\
		\hline
		EMG & Electromyography\\
		\hline
		CT & Computed tomography\\
		\hline
		MRI & Magnetic resonance image \\
		\hline
		WBAN & Wireless body area network\\
		\hline
		BLE & Bluetooth low energy\\
		\hline
		MC & Molecular communication\\
		\hline
		TI & Tactile Internet\\
		\hline
		ML & Machine learning\\
		\hline
		VR & Virtual reality\\
		\hline
		AR & Augmented reality\\
		\hline
		MR & Mixed reality \\
		\hline
		XR & Extended reality \\
		\hline
		MEC & Multi-access edge computing\\
		\hline
		AI & Artificial intelligence\\
		\hline
		DL & Deep learning\\
		\hline
		FL & Federated learning \\
		\hline
		SVM & Support vector machine\\
	\hline
		CNN & Convolutional neural network\\
	\hline
		DNN & Deep neural network \\
	\hline
		GAN & Generative adversarial network\\
	\hline
		GNN & Graph neural network\\
	\hline
		KNN & k-nearest neighbors\\
	\hline
	RL & Reinforcement learning\\
	\hline
	DRL & Deep reinforcement learning\\
	\hline
	LSTM & Long-short-term memory\\
	\hline
	
	\end{tabular}
\end{table}

\section{A Comprehensive Exploration of HDT} \label{SE2} 

\subsection{Differences Between HDT and Conventional DT}
HDT focuses on virtual replicas of human beings and possess unique characteristics compared to the conventional DT (which is mostly applied in industries where physical entities are usually machines). First, human beings are living entities, and the most significant difference between human beings and machines is emotion and psychology \cite{169, 475}. External factors or physiological states can affect individual emotions and psychology, which in turn affect individual physiological states. Second, human's external behaviours depend on individual subjective consciousness, while internal behaviours (e.g., blood flow, the progress of diseases, activities of organs and tissues) are known to generally result from multi-source and complex factors. On the contrary, the behavioural rules of machines are similar and predetermined. As a result, humans are particularly complicated systems with more uncertainty compared to machines, and the abstract process of humans is significantly more difficult than machines \cite{169, 475}.

On top of this, ethical consideration is another unique feature of HDT \cite{221, 475, 169}. This may potentially lead to healthcare inequality between the developed and developing countries. Additionally, since HDT can reflect a patient's real-time and accurate health status, the question of whether or not the patient has the authority to access the actual health status when the patient is diagnosed with a certain disorder needs to be considered on the ethical level.

Furthermore, human entities' data are more heterogeneous and unstructured. As a result, multi-source and sophisticated data are required to provide a high-fidelity digital representation model of any human entity. Asides from physiological data, the commonly unstructured environmental and social media data are important when abstracting human virtual twins because of the high correlation that exists between humans and such external data \cite{169, 475}.
Lastly, humans are mobile agents, unlike machines. This human mobility poses several challenges to the design of HDT, making the migration and placement problems of HDT important issues to address \cite{387}.

The distinctions between HDT and the conventional DT are summarized in Table \ref{table1}. 

 \begin{table*}[!ht] 
	 	\centering
	 	\caption{Differences Between HDT and the Conventional DT}
	 	\label{table1}
	 	\begin{tabular}{| m{3cm} <{\centering} | m{7cm}  <{\centering}| m{7cm} <{\centering} |}
		 		\hline
		 		\rowcolor{lightgray}\textbf{Key Features} & \textbf{Conventional DT} & \textbf{HDT} \\
		 		\hline
		 		Emotion and Psychology & Lack of emotions and psychology. & Human beings are living entities. Hence, their emotions and psychology can be affected by external factors or physiological states. \\
		 		\hline
		 		Behavioural Rules & The behavioural rules of similar machines are almost predetermined and similar. & Human external behaviours depend on individual subjective consciousness and internal behaviours, which are affected by multi-source and complex factors. \\
		 		\hline
		 		Ethical Consideration & Limited or no ethical considerations. & HDT can lead to healthcare inequality between developed and developing countries. Besides, it is not clear whether patients should be authorized to access some information about their own health conditions.   \\
		 		\hline 
		 		Data Complexity & Mostly structured and homogeneous.  & Mostly heterogeneous and unstructured. Correlation also exists between each individual and external data such as environmental and social media data.\\
		 		\hline
		 		Mobility & Mostly fixed with no or limited mobility & Highly mobile with very complicated mobility patterns. \\
		 		\hline
		 		
		 	\end{tabular}
	 \end{table*}

\subsection{Traditional Digital Twin in Healthcare is Evolving Toward HDT}

\begin{figure*}[!t]
	\centering
	\includegraphics[width=0.9\textwidth]{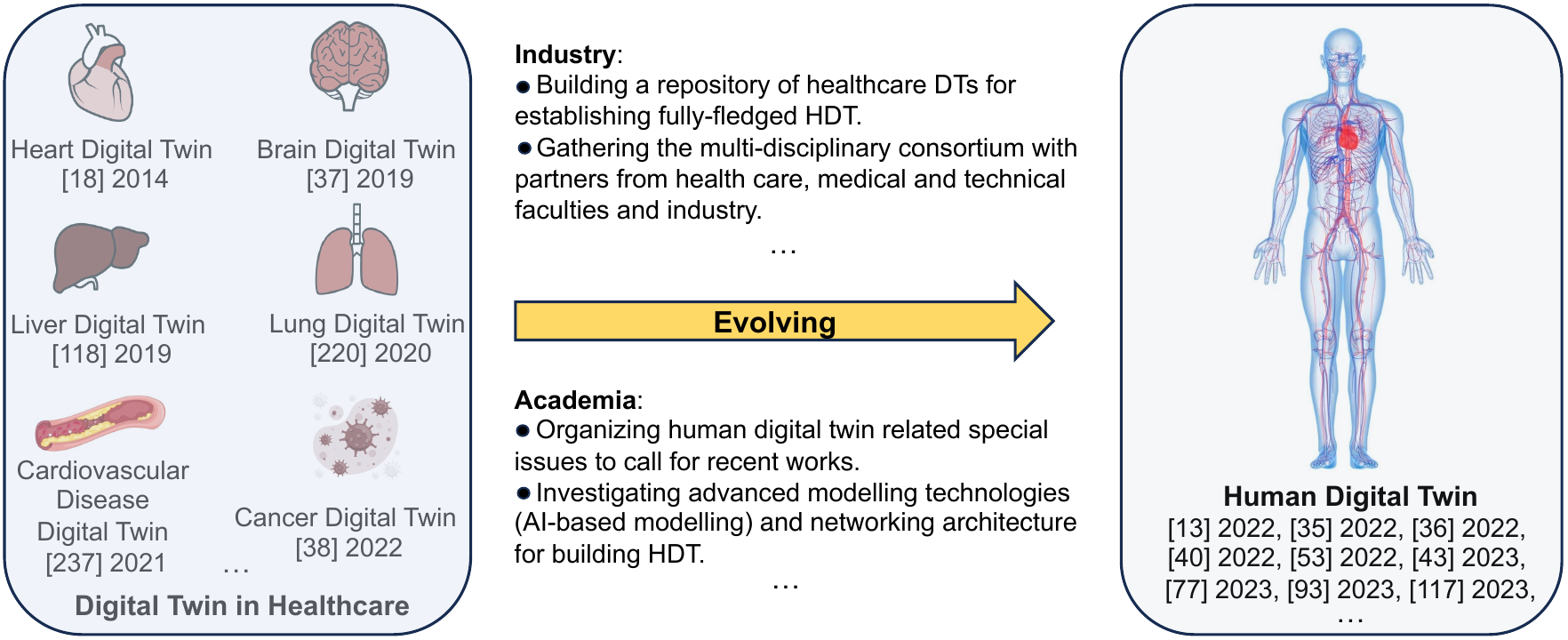} \\
	\caption{DT in healthcare is evolving toward HDT.}\label{DTHtoHDT}
\end{figure*}

The research of the traditional DT in healthcare can be dated back to 2014, when the Dassault Syst\`{e}ms released the SIMULIA living heart project \cite{34}. The project developed the world first DT of human heart that replicates the same functions as a physical heart. This project has proven that DT technology can act as a significant role in cardiac disease research and treatment, and has inspired the researchers from various areas to explore the great potential of DT in healthcare. Since then, this area has attracted a myriad of attentions from both academia and industry. For instance, Neuroelectrics Barcelona held an EU-funded project, called Neurotwin \cite{528}, to develop personalized brain models based on neuroimaging data from Alzheimer’s patients, looking for ways to restore healthy brain dynamics. Batch et al. in \cite{529} developed a cancer DT to improve the detection of metastatic disease. Obviously, the DT in healthcare focuses more on constructing a DT for a particular human organ, cell or disease, which is limited to partial functions of the human body, and cannot comprehensively characterize the complete life cycle of human. However, it is indeed the foundation of realizing HDT.

As recently recognized in \cite{387, 475}, HDT is expected to be an exceptionally vivid in silico representation of human profile for broad and refined applications, ranging from health monitoring, diagnosis, prescription, clinical trials, surgery, rehabilitation to the others, in dealing with various diseases, providing more effective and personalized healthcare services. As shown in Fig. \ref{DTHtoHDT}, HDT is a more holistic and versatile concept that the traditional DT in healthcare is continually evolving towards to, along with advancements in key supporting technologies, such as the modelling technology (e.g., AI-based modeling), computation hardware (e.g., high-performance graphics processing unit) and others. For instance, European Ecosystem for Digital Twins in Healthcare (EDITH) project, which was launched in 2022, has built an ecosystem, where all members of the consortium aim to develop a repository of healthcare DTs, and work collaboratively to achieve the ultimate objective, i.e., the establishment of a fully-fledged HDT \cite{530}. Besides, Ceduersund et al. in \cite{532} have developed sophisticated DTs of all main organs of the human body (including the adipose and muscle tissue, liver, brain, pancreas, blood, etc.), and have begun interconnecting these individual organ DTs to create a comprehensive HDT that can be utilized for various applications.

All these indicate that, with the increasing popularity of this research direction, people have reached a consensus to name this concept as HDT, i.e., a more holistic and versatile concept that the conventional ``DT in healthcare'' can serve as the foundation but eventually evolves towards to.

\subsection{The Universal Framework of HDT}
\begin{figure*}[!t]
	\centering
	\includegraphics[width=0.99\textwidth]{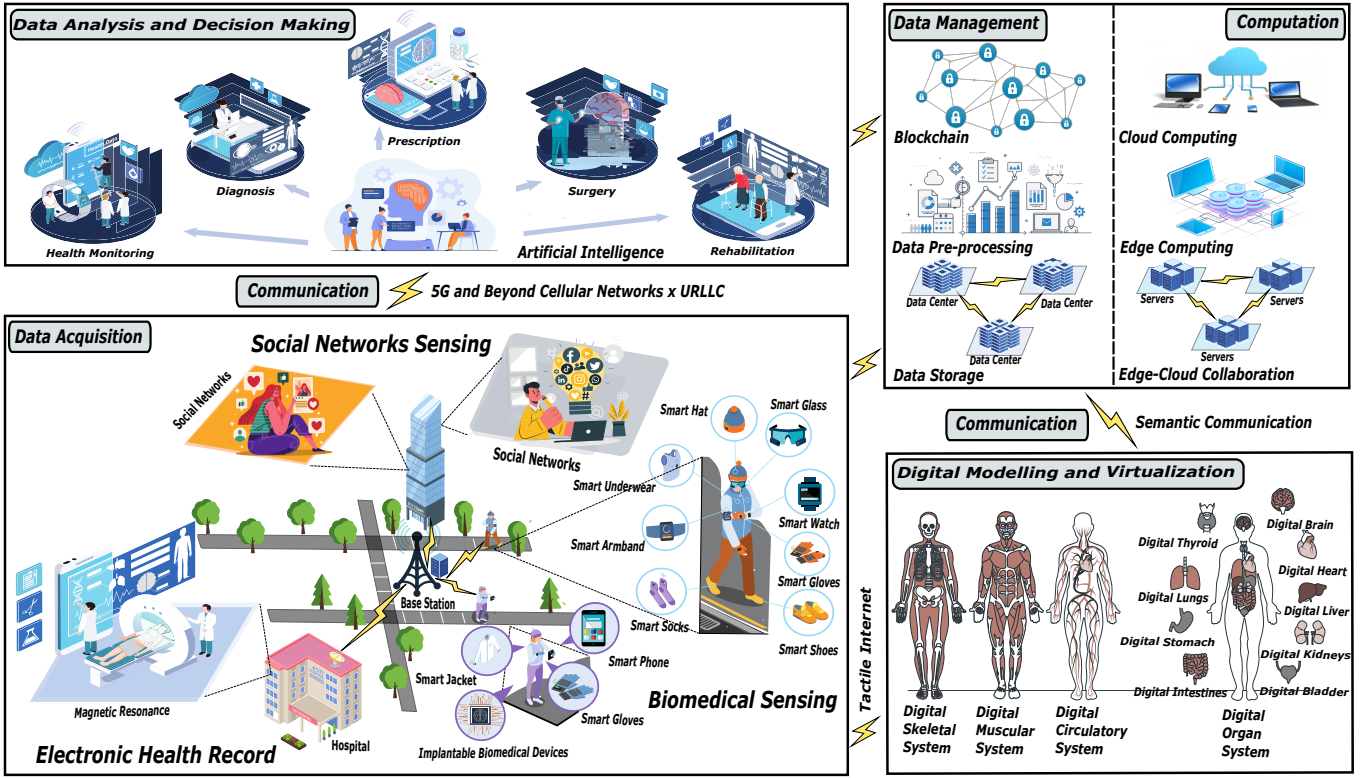} \\
	\caption{The universal framework of HDT.}\label{HDT_arc}
\end{figure*}
HDT is capable of revolutionizing the current healthcare sector. As shown in Fig. \ref{HDT_arc}, we summarize the universal framework of HDT from \cite{387, 476, 477, 478}, which consists of six fundamental components, i.e., \emph{data acquisition; digital modelling and virtualization; communication; computation; data management; and data analysis and decision making}. 


\subsubsection{Data Acquisition}
 Since HDT is a data-driven model, a reliable data acquisition process is vital for HDT. HDT requires both physiological and psychological data of each physical twin (PT), possibly acquired through multiple sources, to establish high-fidelity digital representation of virtual twin (VT). Specifically, it includes medical data from medical institutions, such as electronic health records (EHR) including biomedical examinations and medical images, physiological data (e.g., heart rate, blood pressure and the concentration of biomarkers) obtained through smart personal medical devices, and data from users' social media, such as posts, comments or messages on Facebook, Instagram and Twitter, which can be used to estimate emotions and psychologies.
\subsubsection{Digital Modelling and Virtualization} 
 In HDT, a VT, built on \emph{in silico}, is a virtual replica of its corresponding PT located in the physical environment and co-evolutes with such a PT via reliable connections. By adopting various digitization technologies, physical geometries, properties, behaviours and rules of each PT are digitized holistically to create high-fidelity VT. Such VT depends on real-world data from the physical world to formulate human real-time status \cite{510, 318}. After the digital modelling process, authorized users such as PT, caregivers, relatives and medical personnel can access the VT through interaction technologies, such as tangible XR and hologram, to have immersive interactive experiences with VT.
\subsubsection{Communication and Computation}	
 Communication and computation are significant for HDT. Communication schemes facilitate real-time connectivity among PTs and VTs to ensure synchronization. These connectivity schemes include PT-VT, VT-PT, PT-PT and VT-VT connectivity modelling. Similarly, computation schemes are required in HDT for the execution of various tasks. Data must be properly extracted, processed, securely transmitted and executed through some AI-driven techniques.
\subsubsection{Data Management} 
 Since data are obtained through multiple sources including related PTs and VTs, the size of data in HDT is massive. These data are generally heterogeneous, multi-scale, multi-source and with high noises. Therefore, HDT requires efficient and effective data management frameworks to ensure the construction and evolution of VT.
\subsubsection{Data Analysis and Decision Making}
This component enables HDT to provide reliable data-driven analytics, with the ability to accurately extract underlying information and knowledge from any received massive data, thereby enhancing PH services.


\subsection{HDT: A Hyper-Realistic and Hyper-Intelligent Testbed}
\subsubsection{Profoundly Immersive Experience}In HDT, a high-fidelity VT is built using powerful digital modeling and virtualization techniques with real-time information collected from the corresponding PT. Based on this, users access to a VT with immersive equipment can obtained profoundly immersive experience. For example, doctors utilize the VT of a patient to establish a personalized surgical procedures before the actual surgery. In this scenario, doctors with VR and tactile equipment, among others, interact with the VT, and all the human sensations, such as haptics (e.g., sense of touching skin) and visual (e.g., flow of blood) can be fed back vividly and timely to the doctors' equipment.
\subsubsection{Interaction-Driven Optimization}HDT is expected to be a hyper-intelligent human body testbed for optimizing physical fitness. For example, a doctor could use HDT to prescribe medication by testing various potential prescriptions on the patient's VT, which would save costs and result in a more personalized prescription. The virtual prognosis accelerated by the powerful computing power will output much quicker and potentially more accurate compared to the physical world. The results feedback to the doctors, and the doctors would then optimize the prescription based on such feedback until the optimal prescription is achieved. In addition, the VT with real-time information collected from its corresponding PT could analyze or even predict the PT's health status, and provide timely healthcare recommendations for the PT.

Overall, as one of the major features of HDT, feedback commonly carries massive multimodal information, whether in profoundly immersive experience or interaction-driven optimization scenarios. Achieving ultra-low round-trip time between physical and digital spaces requires significantly efficient network resource optimization. These requirements will be discussed in detail in the next subsection.

\subsection{Design Requirements and Challenges}
It is worth noting that HDT is a complex system with many correlated components, presenting many similarities to conventional DTs. Especially when considering specific use cases of HDT in various PH applications, such as real-time healthcare monitoring, personalized diagnosis and personalized prescription, it is obvious that such application scenarios pose a number of stringent design requirements and challenges for HDT, as discussed below.
\subsubsection{Sophisticated and High-Quality Data}
    Data are essential for HDT. The data for HDT should be sophisticated, which means that the data should be large-scale, real-time, multi-source and multi-modal, and possess deep values. Specifically, HDT needs massive data gathering from multi-source to build a high-fidelity VT of a human. The multi-source and multi-modal data involves not only human data, but also environmental data, providing more accurate information for the construction of HDT through mutual supports, supplements and corrections, for satisfying different HDT requirements. Based on this, the VT in the digital space needs real-time data to timely update itself to keep synchronized with the PT. Besides, HDT needs data with deep values, where HDT could gain deep insights from the collected data for providing accurate and forward-looking feedback.
	However, missing or inaccurate data poses a serious risk to the evolution of HDT and can often lead to misleading information and suggestions from VTs. Such scenarios are undesirable in HDT systems since the outcome can be disastrous, thereby undermining the essence of HDT. Therefore, to ensure that each VT is an accurate replica of its PT, high-quality and noise-free data must be also effectively shared among each PT-VT pair for appropriate model evolution and subsequent decision making. 

	Although some routine healthcare data (such as step number, heart rate and body temperature) and medical data (e.g., medical images) can be easily captured by common sensing devices such as accelerators, gyroscopes, pressure sensors or through manual processes, there exist many physical states that are difficult to capture. For instance, irritation measured through skin rubbing count and polyphagia measured through food intake count, which are two essential health insights when predicting the risk of diabetes \cite{52}, are difficult to be captured in physical states. Therefore, specific biomedical sensors must be designed to retrieve these information.
	Additionally, in HDT, since data collection frequencies of various data are inherently different, and the data may be collected from multiple sources, asynchronous data acquisition and multimodal data fusion are required to be addressed when establishing HDT.
\subsubsection{Extreme Ultra-Reliable and Low-Latency Communication (xURLLC)}
	PT and VT will generate and exchange high volume and multidimensional data to maintain synchronization. This synchronization is expected to be supported by xURLLC, with data transmission rate of $\geq$ 100 Gbps, reliability of $\geq$ 99.99999$\%$ and latency of $\leq$ 1 ms \cite{50}. Specifically, when adopting HDT in time-critical healthcare applications, e.g., remote surgery \cite{231}, xURLLC is required to support real-time update of the VT as well as timely receptions of feedbacks from such a VT to facilitate timely decision optimizations at the paired PT.
	Furthermore, important interaction technologies such as tangible XR \cite{343} --  a technology that combines XR and tactile internet to transmit not only large capacity contents such as video and three-dimensional computer graphics but also haptic signals, such as feelings of touch – also requires the supports of xURLLC \cite{230}.
	However, current networking technologies, i.e., fifth generation (5G) mobile networks, characterized by ultra-reliable and low-latency communications (URLLC), cannot meet these stringent requirements. Therefore, future communication technologies are necessary to provide xURLLC services to support HDT applications. 
	Nevertheless, xURLLC is still inevitable to the high signaling overhead (e.g., clock synchronization and handover costs), which may considerably deteriorate the network efficiency in general, leading to the demand of further integrating the deterministic network technologies, e.g., time-sensitive networking (TSN) \cite{466}.
	
\subsubsection{Ultra-Low Round-Trip Time (RTT)}
In specific applications, e.g., surgery simulation and massage simulation, is imperative to support immersive interactions. For example, one of the most significant challenges in haptic communication is achieving an RTT of 1 ms \cite{474}. RTT is great affected by queuing and processing delays at the intermediate nodes and packet transmission time. However, the packet size transmitted between PTs and VTs are typically large, involving massive multimodal information (e.g., text, audio, video, image, and haptic) to achieve a profoundly immersive experience, which poses a great challenge for achieving ultra-low RTT. Migrating the VT to the vicinity of users (e.g., the corresponding PT or doctors) can be a promising solution. Nevertheless, this will trigger other issues, such as the deployment of migrated VT and timely synchronization updates between the VT and its distant PT. Overall, guaranteeing the ultra-low RTT is essential in HDT applications, and novel network traffic scheduling schemes for HDT may be developed with a careful consideration of this metric.
\subsubsection{Data Privacy, Security and Integrity}
	Healthcare-related data with individual private information/metadata (e.g., name and address) are privacy-sensitive and have no or limited tolerance for privacy leakages \cite{507, 508}. Appropriate mechanisms must be developed to ensure that such data are not intercepted or modified by unauthorized users while being stored and transmitted over the network. 
	In other words, key technologies should provide secure and reliable communications among PTs and VTs with sufficient data storage privacy. Authentication of data sources is also a necessity to ensure that all sources are reliable, while fake data are detected and removed through reliable security measures to guarantee integrity.  
\subsubsection{Data Storage}	
	HDT relies on multi-source real-time data from the physical world. Each HDT application can generate up to a few gigabytes of data in a single day. 
	Storage of such massive data may be required for VT update process and analytics. Therefore, key technologies are required to provide appropriate mechanisms to store the huge amount of data.
\subsubsection{Advanced Computing Power}	
	In HDT, tasks such as synchronization, model evolution and analytics are expected to be time-sensitive and computation-intensive. To ensure a real-time computation process, massive computation resources are required by HDT to keep high-fidelity. This prompts HDT to be deployed on the network side instead of local devices (which are commonly resource-limited). Furthermore, this paradigm requires optimal computation resource scheduling mechanisms on the network side to guarantee the efficient resource utilization.
\subsubsection{AI-Driven Analytics}	
	AI is a core key technology for enabling HDT to have the ability to deliver PH services. 
	AI-driven analytical models provide insights at different scales for HDT using real-time and historical collected data, thereby providing decisions or predictions to individuals and updating the VT model. Additionally, AI can support HDT from all aspects (e.g., intelligent computation, intelligent communication network and intelligent data management). However, one limitation of current AI solutions is that they rely on black-box models and may be inappropriate for problems requiring explicit explanations (e.g., clinical applications). Therefore, the interpretability of AI is an imperative issue that extremely depends on the implementability of HDT. Moreover, the ever-changing of human conditions impel AI models to dynamically validate their effectiveness and update for more precisely abstracting the real status of the human body. However, this process is highly complex, which requires the assistance of a strong computation capability.

\subsection{Overview of Networking Architecture and Key Supporting Technologies for HDT}
\begin{figure}[!t]
	\centering
	\includegraphics[width=0.99\columnwidth]{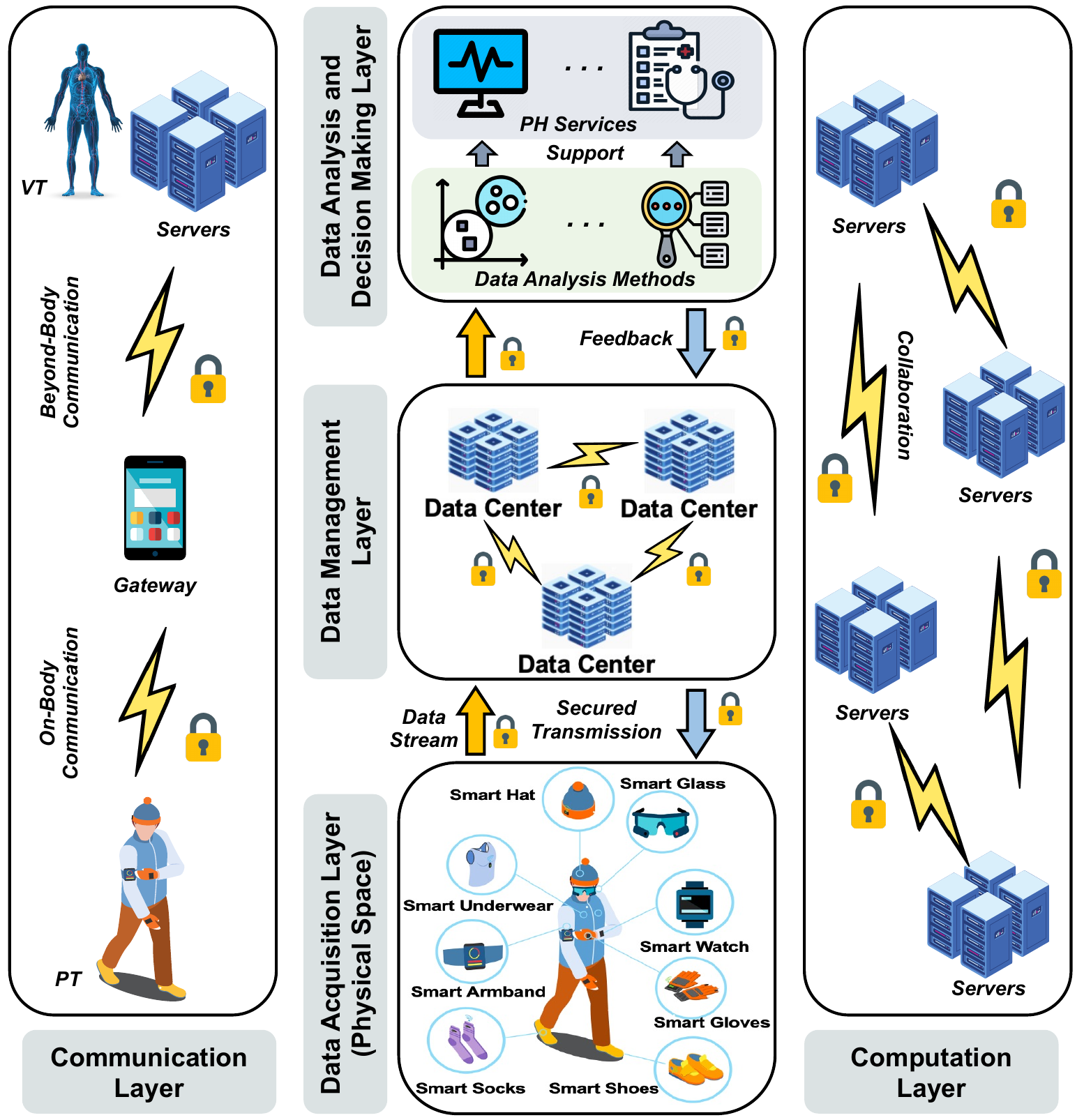} \\
	\caption{The networking architecture of HDT.}\label{Net_Ar}
\end{figure}

In summary, a networking architecture is imperative to enable the realization of HDT for PH applications. As shown in Fig. \ref{Net_Ar}, it is expected that the networking architecture consists of five layers according to the end-to-end data stream processing procedure. The data are first collected by the data acquisition layer, and are transmitted to the data management layer through the communication layer for pre-processing, storing and sharing, and then by the support of the computation layer, they are processed by the data analysis and decision making layer for serving powerful applications (e.g., diagnosis, prescription, surgery and rehabilitation). In return, the feedback of the VT in the digital space to its corresponding PT is transmitted through all these five layers reversely to the physical space.

Specifically, since HDT is an extremely complex data-driven system, it needs multiple and heterogeneous nodes to collaboratively acquire sophisticated and high-quality data in the data acquisition layer (similar to the structure of sensor networks). Then, for supporting the heterogeneous data transmission requirements in HDT, multiple heterogeneous communication paradigms are needed in the communication layer, which includes communication protocols on and beyond human bodies. Since PTs in the physical environment is mobile, while requiring fast-responsive and muscular computing power services for supporting time-sensitive and computation-intensive tasks in HDT (e.g., model evolution, real-time rendering), multi-node collaborative computing paradigm with frequent information exchange is required to provide ubiquitously advanced computing power in the computation layer. Furthermore, considering that the data generated in HDT (e.g., data from data acquisition layer and feedback data from VTs) are typically large-scale, multi-modal, multi-source, and with high noises, it requires multiple servers to collaboratively pre-process and store the data for providing the robust data management service in the data management layer (similar to the structure of data center networks). On top of this, data security and privacy can also be potentially offered by such paradigm through multi-node collaboratively authentication, among others. For simultaneously realizing HDT's different PH services (e.g., synchronization of the PT and VT pair, health monitoring and diagnosis), multi-heterogeneous data analysis methods are expected to play vital roles in the data analysis and decision making layer. Overall, the networking architecture of HDT is a complex and sophisticated system that requires collaboration across multiple layers and nodes.

\begin{figure*}[!t]
	\centering
	\includegraphics[width=0.95\textwidth]{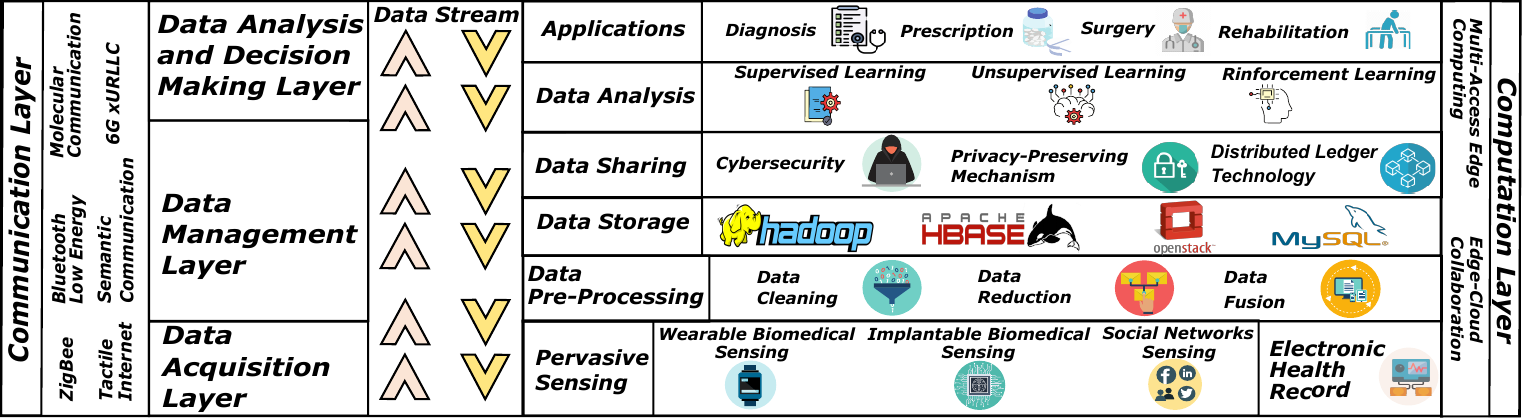} \\
	\caption{An overview of key technologies for implementing the networking architecture of HDT. }\label{Net_tec}
\end{figure*}

For comprehensively implement such networking architecture, a variety of key end-to-end technologies are required, as shown in Fig. \ref{Net_tec}. Particularly, in the data acquisition layer, sensing devices such as wearable biomedical devices and implantable biomedical devices can be adopted to enable pervasive sensing. In addition, social networks and electronic health records can also serve as data sources to abstract the physiology and psychological status of any PT. 
	In the data management layer, data cleaning, data reduction and data fusion technologies can be used to pre-process the data generated in HDT, before actual utilization and potential storage. The big data storage frameworks (e.g., hadoop distributed file system, HBase and openstack swift) can be tailored to robustly storing HDT data. To guarantee data security and privacy, existing tools of cybersecurity, privacy-preserving mechanisms and distributed ledger technology can be applied. 
	In the data analysis and decision making layer, powerful AI algorithms, such as supervised learning, unsupervised learning and reinforcement learning (RL), can be employed to facilitate the HDT-enabled PH applications including personalized diagnosis, personalized prescription, personalized surgery and personalized rehabilitation.
	In the communication layer, the communication paradigms of HDT must be carefully tailored through modifications of existing and emerging communication techniques, such as Bluetooth, ZigBee, molecular communication, tactile Internet and semantic communication. While the computation layer can be established  by integrating novel computation paradigms, including multi-access edge computing and edge-cloud collaboration.

\section{Key Technologies for Data Acquisition Layer} \label{SE3}
Unlike conventional DT, the data collected in HDT mainly originates from individuals with complex mobility. This includes the acquisition of diverse physiological and psychological data, which are obtained through various biomedical sensors \cite{506}. Further elaboration is provided below.

 Collecting human physiological data for building HDT can be enabled by wearable and implantable devices. Smart wearable devices with biomedical sensors (e.g., smart watches, smart socks and smart garments) are developing rapidly in recent years. These wearable devices offer an exciting opportunity for measuring human physiological signals in a nonintrusive and real-time manner by leveraging flexible electronic packaging and semiconductor technology \cite{458}.
Nevertheless, smart wearable devices are limited to monitoring only specific types of physiological parameters that are readily accessible from outside the human body (e.g., body temperature, heart rate and step number). Smart in-body biomedical devices with implantable biomedical sensors that are placed directly inside human bodies promise an entirely new realm of applications. Besides, the development of nanotechnology and its application in medicine, through nanomaterials and nanodevices, enables in-body biomedical devices to have diverse clinical applications, such as biomarker concentration monitoring, network tomography inference of human tissues and monitoring of oxygen levels in the surrounding tissues. 
 
\begin{figure}[!t]
	\centering
	\includegraphics[width=0.95\columnwidth]{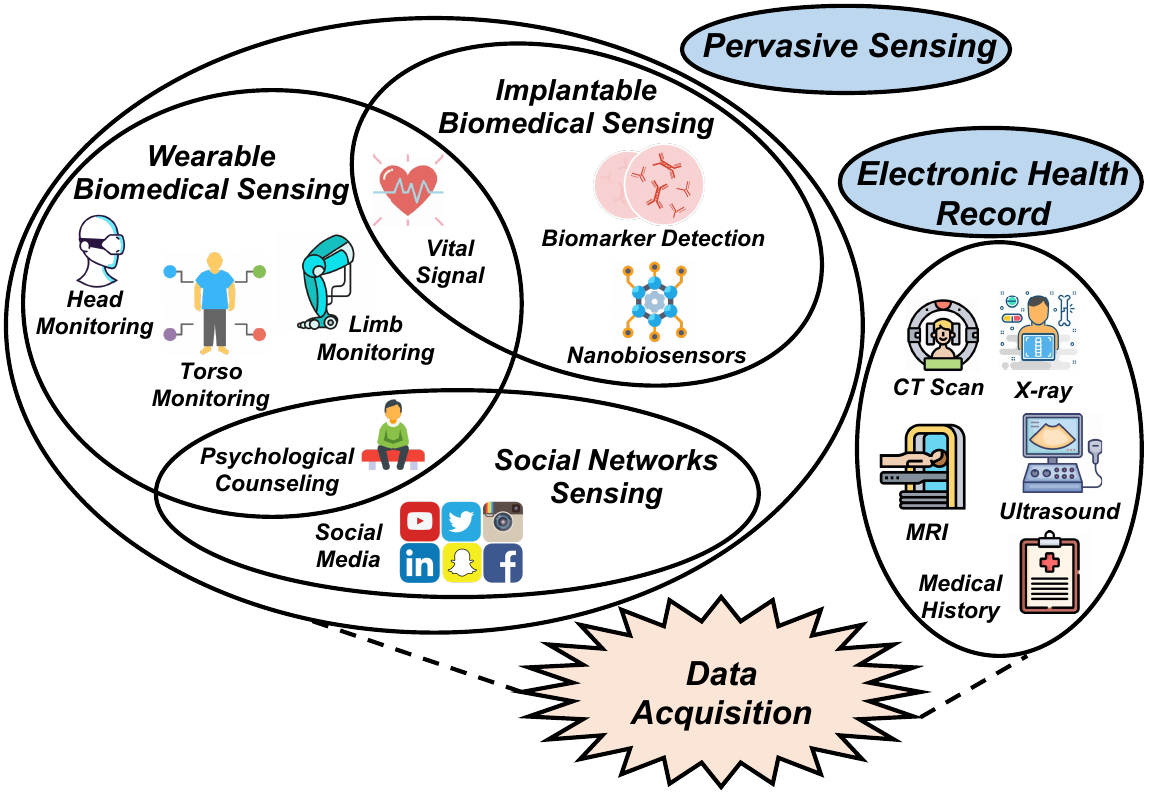} \\
	\caption{An illustration of data acquisition approaches for HDT.}\label{data_acq}
\end{figure}

In addition to physiological data, the high-fidelity digital model of humans also needs psychological data. Software-based soft sensors collect data mainly from social networks, such as Instagram, Twitter and Facebook, where humans sometimes post information about their feelings and emotions \cite{43}. Asides from data obtained through various sensing devices (from pervasive sensing and social networks sensing), electronic health records (EHRs), which record individual health-related data generated by various medical institutions, is another important source of data for HDT applications. All different data acquisition approaches for HDT are visually illustrated in Fig. \ref{data_acq}.

\begin{figure}[!t]
	\centering
	\includegraphics[width=0.99\columnwidth]{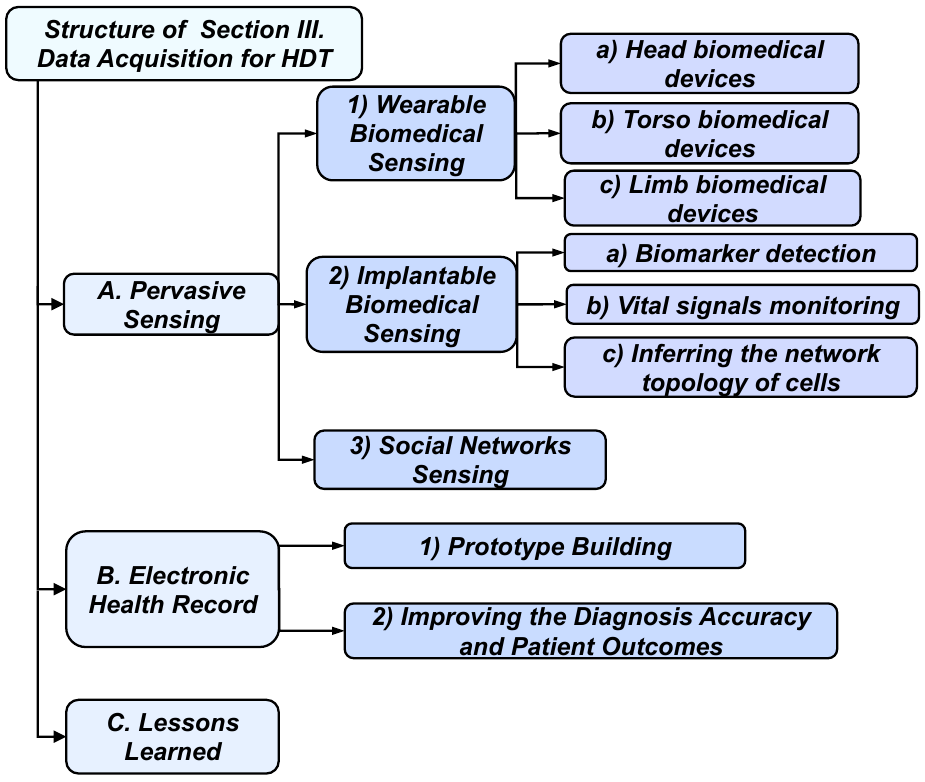} \\
	\caption{The roadmap of Section \ref{SE3}.}\label{Stru_3}
\end{figure}

In this section, we provide a review of data acquisition solutions for HDT. First, we discuss wearable biomedical sensing, implantable biomedical sensing and social network sensing in Section \ref{SE3_1}, which allows HDT to collect real-time physiological and psychological information of PTs. Besides, EHRs is another crucial data source accurately reflecting PTs' health information, which can be used to build the prototype of VTs and improve diagnosis accuracy and patient outcomes as reviewed in Section \ref{SE3_2}. Finally, in Section \ref{SE3_3}, we provide a brief summary of the reviewed papers, and discuss some opening issues that should be considered in data acquisition for HDT. The roadmap of Section \ref{SE3} is illustrated in Fig. \ref{Stru_3}.

\subsection{Pervasive Sensing} \label{SE3_1}
In PH applications, there exists a high-natural variability which can be explained through the innate differences of human bodies, such as disease progression or response to medical treatments. These parameters need to be understood by biomedical sensors implemented in HDT to build a personalized digital representation of a PT and prevent false positives. 
Therefore, in this section, we delve into the Internet of medical things (IoMT) devices, which consists of wearable biomedical devices, implantable biomedical devices embedded with powerful sensors, and social networks that are often implemented in HDT to ensure pervasive sensing.

\subsubsection{Wearable Biomedical Sensing}
Wearable biomedical devices developed so far have been designed for different functions and positions on human bodies. These wearable devices can be classified into three categories, i.e., head, torso and limb wearable biomedical devices. Wearable biomedical devices equipped in those human body parts can sample diverse physiological data which is required by the construction of HDT. We survey them in the following.

\begin{enumerate} [ a) ]
	\item \emph{Head biomedical devices:} As the utmost part of a human body, the head includes various important organs, such as eyes, nose, ears and mouth. Corresponding wearable health-tracking biomedical sensors designed for the head may be smart glasses, contact lenses, helmets, hearing aids, earrings, etc.

	\begin{itemize}
	\item Smart glasses function as a kind of wearable micro-computers embedded along with many biosensors, like gyroscopes, accelerometers (ACC) and pressure sensors. 
	Smart glass are versatile platforms, which can assist in weight management by detecting and recording the eating and drinking habits of individuals \cite{236}, and provide individuals with real-time electrocardiograms (ECG) monitoring\cite{237}, while providing feedback (e.g., healthcare recommendations) to help users (especially people with Parkinson) with self-management \cite{239}, etc.
	\item Smart contact lenses, in terms of the architecture, may consist of multiple biomedical sensors (e.g., capacitive, strain, microfluidic, channel electrochemical, fluorescent and holographic biomedical sensors) \cite{241}. It is similar to implants and can be worn or removed easily by users. Smart contact lenses can be used for physiology monitoring by continuously detecting glucose levels in tears,
	while tracking the progression of patients’ glaucoma by continuously measuring eye lens’ curvature \cite{242}.
	\item Smart helmets are commonly embedded with biosensors such as infrared temperature and heart rate biomedical sensors. A smart helmet can be used as an alternative for monitoring parameters commonly obtained by other smart wearable devices, e.g., body temperature and heart rate, via sensors located inside and around the helmet \cite{243}. In addition to these, helmets worn on the head can also be employed for detecting brain activities. For instance, SmartCap Technologies developed a smart helmet, called SmartCap, working as a fatigue tracking system, which possesses the ability of measuring brainwave signals to alert the potential risk of microsleeps \cite{245}. Besides, electroencephalographic (EEG) headsets can monitor the EEG of brain signals to measure mental activity, such as tracking a user's confusion states for assessing or quantification the user's focus level \cite{246}.
	\item Smart wearable devices worn on ears are usually in the form of hearing aids, earphones and earbuds. These devices possess the ability to monitor physiological parameters such as ECG signals, breathing rate, pondus hydrogenii and lactate values of sweat, using biomedical sensors like amperometric and potentiometric biomedical sensors \cite{247}.

	\item Some wearables can also be worn inside the mouth. Examples of those devices are smart mouth guard (MG)-type wearable devices which can monitor teeth clenching with embedded force sensors \cite{250}. Another example is DentiTrac, which is a miniature oral device integrated on an oral appliance, used for monitoring patients’ sleep apnea and adherence \cite{251}. 
	\end{itemize}

	\item \emph{Torso biomedical devices:} Torso is the central part of the human body, where many vital organs are located. Common examples of wearable devices placed on the torso are smart clothing, belts and underwear. 
	\begin{itemize}
		\item The realization of smart clothing significantly depends on smart textile technologies, where the health monitoring sensors are completely embedded in the fabric. 
		Such clothes can offer comfort and smart healthcare to their users. For instance, a smart jacket integrated with a health monitoring system was developed in \cite{252} to monitor the pulse rate and EEG in the human body.
	
		\item Skin patches/tattoos with biomedical sensors may also have a great potential in continuous monitoring vital physiological signals to improve the healthcare quality of patients. Several key applications of smart patches/tattoos have been demonstrated in ECG monitoring \cite{257}, pulse rate monitoring \cite{259} and biomarker measurements in sweat \cite{258}.
		\item A typical application scenario of smart belts embedded with sensors such as inertial sensor and bend sensor is for monitoring shoulder and trunk posture \cite{ 263}. A smart belt can also be designed for monitoring physiological signals such as real-time respiratory signs monitoring \cite{ 265} and detection of respiratory rate, body movement, in-and-out-of-bed activity and snoring events during sleep \cite{266}.
		
	\end{itemize}
	\item \emph{Limb biomedical devices:} The four limbs of the human body are the main executors of activities. Wearable devices worn on limbs are mostly accessories, such as smart bracelets, watches, armbands, rings and wristbands. These devices can monitor physiological parameters while posing little or no interference to the users' normal activities.
	\begin{itemize}
		\item Smart armbands are wearable devices with embedded sensors such as photoplethysmography (PPG), ACC and ECG sensors, and are usually worn on the upper arms to facilitate seamless health monitoring with maximum comfort to the wearer. Application examples include continuous estimation of the respiration rate \cite{280}, measurement of blood pressure \cite{281} and ECG signals \cite{282}. 
	
		\item Smart pants with embedded sensors can be used collect physiological data. For instance, a pair of pants embedded with an ACC, gyroscopes, and WIFI probes was used in an HDT to detect the patient's physical activity and vital signs, serving as a data source for type 2 diabetes management \cite{512}.
		\item Smart shoes can be utilized for monitoring human physical activities. For example, a pair of shoes equipped with inertial sensors was used to track gait events during stork rehabilitation in an HDT system \cite{522, 534}.
	\end{itemize}
	
\end{enumerate}
\subsubsection{Implantable Biomedical Sensing}
The advances in nanotechnology continue to facilitate the growth of implantable biomedical sensors such as implantable nanobiosensors. 
Unlike wearable devices, implantable biomedical sensors are generally deployed inside human bodies, for instance, on organs and bloodstream, to perform more powerful tasks ranging from precision drug delivery, precision sensing and micro procedures in the inaccessible organs of the body \cite{295}. In this subsection, we focus on the abilities of implantable biomedical sensors to facilitate precision monitoring and activity measurements such as disease biomarker detection, vital signals monitoring and detection of cell network topology.

\begin{enumerate}[ a) ]
	\item \emph{Biomarker detection:} Biomarkers are often released into the blood by certain diseases, such as cancers and diabetes. 
	For instance, isopropanol (IPA) is the biomarker for both types of diabetes \cite{296}, while $\alpha$-Fetoprotein is the common biomarker for hepatocellular carcinoma (HCC) \cite{297}. Continuous monitoring of biomarkers in real-time can significantly advance precision medicine and is a more effective method compared to conventional blood tests, where the concentration of biomarkers in any sample taken is usually very low, especially for chronic diseases in their early stages. When adopted, the biomarker detection technique can aid the detection of the cancer biomarkers concentration in the blood vessels. Through moving the nanobiosensors along the blood vessels of the cardiovascular system, the biomarkers around the cancer cells could be detected, thereby facilitating the early diagnosis of cancer \cite{56, 299, 300}. It can also facilitate continuous measurement of biomarker concentration released by bacterial cells, thereby estimating the number of infectious bacteria while deducing the progress of the infection for early detection of infectious diseases \cite{301}. The biomarker detection can also enhance continuous monitoring of endothelial cells shedding in arteries as an early sign of heart attacks \cite{302}.
	
	\item \emph{Vital signals monitoring:} Asides from biomarker concentrations, implantable nanobiosensors are also useful for continuous monitoring of vital signals including local temperature, cardiovascular system, musculoskeletal system and pondus hydrogenii within the central nervous of the human body. 
	Treatments of traumatic brain injury (TBI) can lead to increased intracranial pressure (ICP), thereby interfering with vital functions. As a result, the ICP must be constantly monitored, for instance through implantable nanobiosensors \cite{309}. Moreover, intracranial temperature (ICT) monitoring is another vital signal since it is associated with changes related to the volume of air inside skulls. The changes in ICT in the range of 35-40 $^{\circ}$C can be monitored by biodegradable intracranial nanobiosensors \cite{310}.
	Besides ICP and ICT, implantable nanobiosensors are also used to monitor intracranial electrical activity -- an important activity when managing neural disorders such as Parkinson's disease, Alzheimer's disease, depression and chronic pain \cite{310}. Furthermore, implantable blood flow monitoring bio-compatible sensors have been developed to wrap around the blood vessels and provide continuous information about the vessel patency \cite{310}. 
	
	
	\item \emph{Inferring the network topology of cells:} Advanced implantable nanobiosensors-based techniques can precisely map cellular connections and allow in-vivo characterization of their activities, thereby assisting the modelling of VT while improving the efficiency of diagnosis of disorders. 
	Indeed, the communications between implantable nanobiosensors rely on the use of tissues as communication channels through the molecular communications paradigm. This makes it possible to send back-back signals between nanobiosensors. These signals can be measured at different points of the tissue to infer the actual network topology of cells. Similarly, the in-vivo cellular activities measurement can be achieved through implantable bio-compatible nanobiosensors. These sensors interface with organs by establishing connections among individual cells to measure cellular signals at intracellular, intercellular and extracellular levels \cite{305}. 
	In \cite{304}, the authors proposed a topology inference technique for human brain cortex neuronal networks based on network tomography theory. An implantable nanobiosensors technology was used to achieve high-resolution and high-precision brain neuron network mapping as well as characterization of neuronal activities.
\end{enumerate}

\subsubsection{Social Networks Sensing}
Different from conventional DT, humans are living entities with complex consciousness, emotions, and psychology. Therefore, the modelling of a high-fidelity digital counterpart (i.e., the VT) must take into account and synchronize with these factors of the corresponding human (i.e., the PT) \cite{513, 517}.
To maintain a typical digital counterpart (i.e., the VT) with high fidelity, both physiological states and emotions of the corresponding human (i.e., PT) must be synchronized in real-time. 
While EEG signals, usually obtained through hard sensors or devices, are common data sources for human emotions recognition \cite{321}, social networks or platforms can similarly play a crucial role in the detection of human emotions \cite{507}.

Nowadays, information including thoughts, mental states and moments of individuals are sometimes available on their respective social network platforms, which can contribute to the amount of psychology-related data being generated every second. For instance, people often share their thoughts and feelings regarding the COVID-19 pandemic or other common diseases through their social network platforms. This data can be processed in real-time to comprehend human current psychological state through the use of sentiment analysis and emotion detection \cite{327}. 
By leveraging the COVID-19 pandemic-related data generated through social network platforms, the authors in \cite{325} carried out sentiment analysis and emotion detection of Twitter users to limit the possibility of various mental health issues such as depression. This effort further justifies the importance of social networks sensing to maintain an accurate synchronization of VT in HDT.

\subsection{Electronic Health Record}\label{SE3_2}
EHRs are real-time, patient-centred records consisting of medical imaging including computed tomography (CT) scan, X-ray, magnetic resonance images (MRI) and ultrasound. It also contains medical and treatment histories, allergies, diagnoses, etc. \cite{314}. EHR can be adopted in HDT to improve diagnosis accuracy and patient outcomes.

\subsubsection{Prototype Building}
EHR data (e.g., medical imaging) can be used to build a prototype VT of a PT (e.g., human, tissues, organs) for personalized healthcare applications (e.g., investigating patients' disease progression \cite{520}).
For instance, a modelling approach for a patient-specific coronary artery (a VT of a coronary artery) was proposed in \cite{315}. The arterial models were developed based on patient-specific medical imaging (i.e., coronary optical coherence tomography (OCT) and angiography) that were acquired during pre-treatment. Tai et al. in \cite{316} reconstructed a 3D patient-specific human lung VT model based on CT images. Ahmadian et al. in \cite{177} created a VT of the human vertebra by relying on a deep convolutional generative adversarial network (DCGAN), which was trained through a set of quantitative micro-computed tomography (micro-QCT) images of the trabecular bone. Gillette et al. in \cite{319} proposed a framework for the generation of the cardiac VT of human electrophysiology. Their solution targeted at providing a digital replica of the human heart using clinically-attained MRI.

\subsubsection{Improving the Diagnosis Accuracy and Patient Outcomes}
The use of EHR data when adopting HDT can improve the diagnosis accuracy and patient outcomes. 
For instance, Allen et al. in \cite{138} used a VT model to forecast the progression of relevant clinical measurements in the patient at risk of ischemic stroke based on EHR data. The results show that this VT model can accurately forecast the disease progression, thereby allowing for tailored treatment to improve patient outcomes. Guo et al. in \cite{320} predicted disease onset information based on the patient's EHR data to guide disease prevention and treatment personalization.

\subsection{Lessons Learned}\label{SE3_3}
Data acquisition in HDT requires massive diverse devices to collect physiological and psychological data for precisely mapping the PT to its digital representation, i.e., VT. 
More specifically, wearable biomedical devices, including head, torso, and limb biomedical devices, are used to measure human physiological signals, while implantable biomedical devices are implanted inside the body to collect physiological data. Social networks can also serve as soft sensors to gather psychology-related data, and electronic health records can be another source of data for HDT, providing information on the patient's treatment histories, allergies, diagnoses and more.
 
 While the diverse data sources mentioned above can be utilized to dynamically map a real-time human body into the digital space, there are several opening issues that need to be considered. 
 First, while human body sensing technologies are continuously advancing, they are still unable to collect ultra-high fine-grained data of the human body.
  Second, to dynamically map the human body, real-time data collection is required. However, it is not feasible to expect humans to wear wearable biomedical devices constantly, and even the battery capacity of these devices, whether wearable or implantable, is limited. Third, integrating those heterogeneous data into a unified model that represents a human body pose a significant challenge. Fourth, the management and storage of such massive data is also challenging. Finally, security and privacy issues of those data have to be well addressed by taking into account ethics and moralities, particularly for healthcare-related data that are highly sensitive.

\section{Key Technologies for Communication Layer} \label{SE4}
\begin{figure*}[!t]
	\centering
	\includegraphics[width=0.95\textwidth]{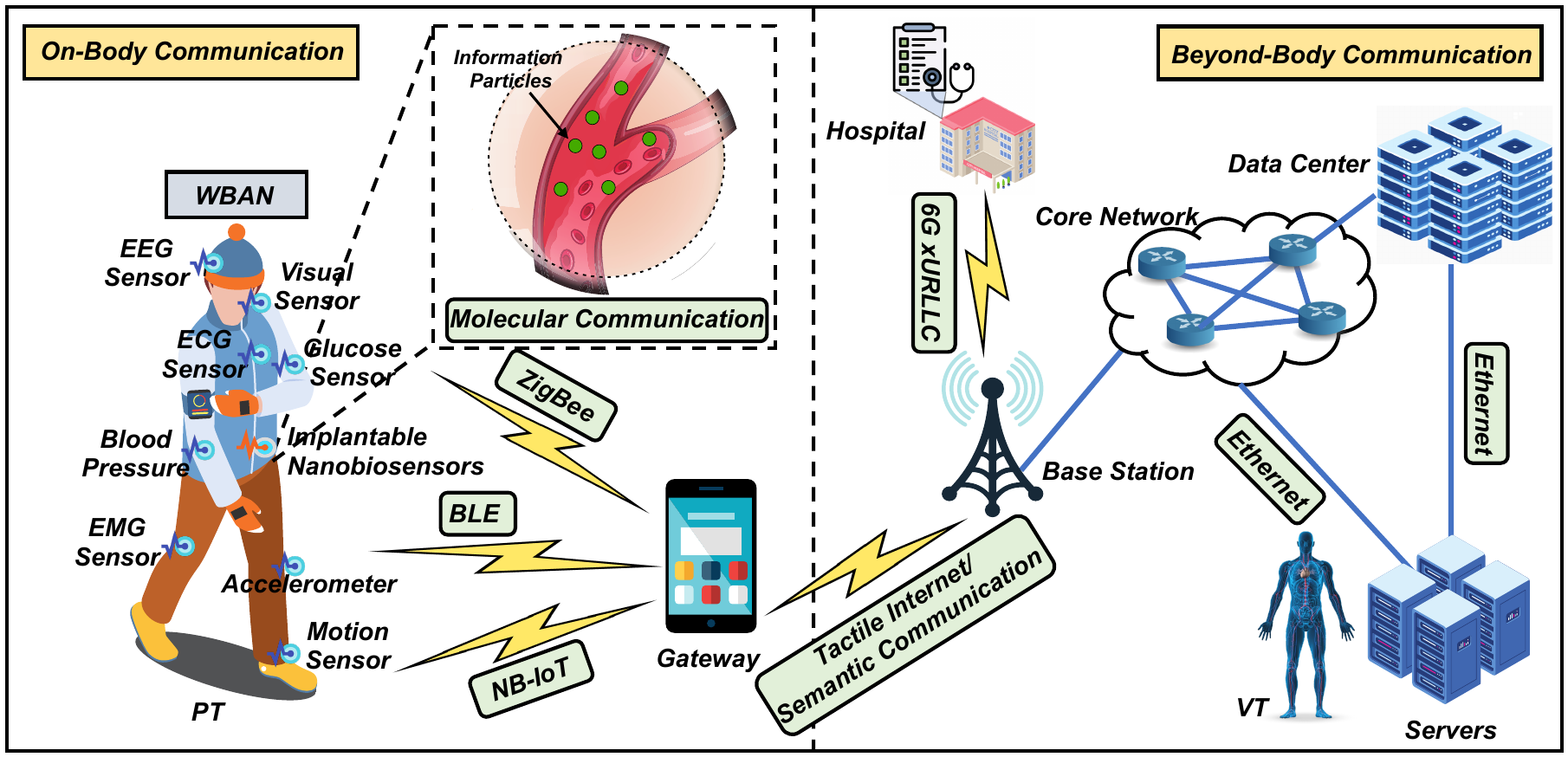} \\
	\caption{The communication architecture of HDT.}\label{com_arch}
\end{figure*}
Communications in HDT can be categorized into two tiers, i.e., on-body and beyond-body communications. On-body communication refers to short-range communications around the human body, typically including communications among body sensors and communications between body sensors and the gateway (e.g., smartphone). Beyond-body communication focuses on communications between gateways and remote servers that host HDT (e.g., cloud servers and data centers). The communication architecture of HDT is demonstrated in Fig. \ref{com_arch} and analyzed in the following subsections.

In this section, we first review on-body communication techniques for collecting physiological and psychological information from PTs to gateways, including Bluetooth low-energy, ZigBee and molecular communication in Section \ref{SE4_1}. Second, in Section \ref{SE4_2}, we discuss that the data transmitted between the physical world and digital world using beyond-body communication usually carries massive and multimodal data for synchronization updates and immersive experience, among others. This can be enabled by tactile Internet for not only transmitting regular information (e.g., text, image and video), also feeding back human sensations (e.g., haptic feelings). Additionally, the explosive growth of data in HDT and limited bandwidth necessitates a paradigm shift away from the conventional focus of classical information theory. In response, we review semantic communication solutions for beyond-body communication of HDT, which can serve to alleviate the spectrum scarcity for HDT applications. Finally, we summarize the reviewed papers, and discuss some opening issues that should be considered in Section \ref{SE4_3}.

\subsection{On-Body Communication}\label{SE4_1}
\begin{figure}[!t]
	\centering
	\includegraphics[width=0.95\columnwidth]{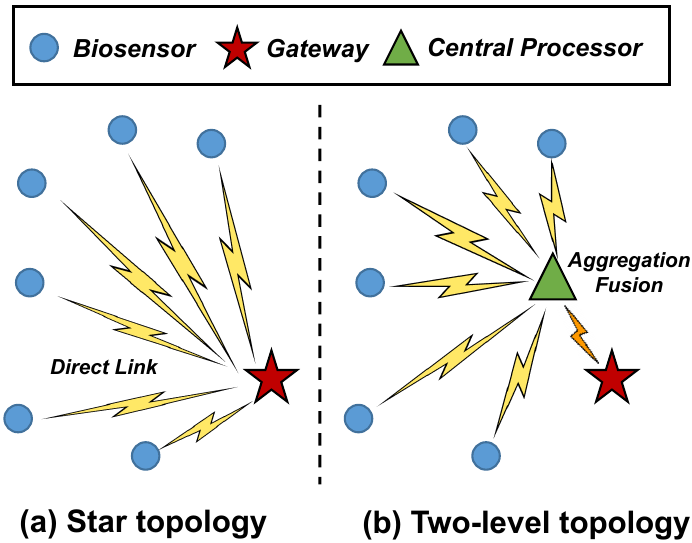} \\
	\caption{The topology of on-body communication.}\label{stru}
\end{figure}
The HDT on-body communication is generally realized through an essential component, called the wireless body area network (WBAN) \cite{456, 482, 484}, where on or in-body sensors are connected and are responsible for the transmission of collected data to the gateway through WBAN. The structure of on-body communication can be divided into two types. In the first type, sensors directly communicate with the gateway, forming a star topology, as shown in Fig. \ref{stru} (a). In the second type, sensors are connected to a body's central processor in the first level for preprocessing to reduce the amount of raw data while saving energy, and then the preprocessed data are forwarded to a gateway in the second level, and thus forming a two-level communication topology, as shown in Fig. \ref{stru} (b). 

Since body sensors are typically low-power, resource-constrained and low bit-rate, they require an energy-efficient and low-range wireless link. Table \ref{table4} exemplifies the information of some human body physiological parameters as well as corresponding data rates. This defines baseline requirements for wireless connectivity. Given the above requirements, the majority of existing implementations in healthcare applications rely on Bluetooth low energy (BLE) or ZigBee to wirelessly transmit the data collected by the sensors to a gateway.

 \begin{table*}[!ht] 
	\centering
	\caption{Information of Physiological Parameters Sampled by Biomedical Sensors (Compiled from \cite{76})}
	\label{table4}
	\begin{tabular}{| m{3cm} <{\centering}| m{3cm} <{\centering} | m{3cm} <{\centering} | m{3cm} <{\centering} | m{3cm} <{\centering} |}
		\hline
		\rowcolor{lightgray}\textbf{Signals} & \textbf{Data Range} & \textbf{Data Rate} & \textbf{Resolution (bits)}  & \textbf{Frequency (Hz)} \\ \hline
		Glucose Concentration & 0-20 mM &  480-1600 bps& 12-16 & 0-50 \\ \hline
		Blood Flow & 1-300 ml/s & 480 bps & 12 & 40  \\ \hline
		ECG & 0.5-4 mV &6-48 Kbps & 12-16 &  0-1000  \\ \hline
		Blood pH & 6.8-7.8 pH units & 48 bps& 12 & 4  \\ \hline
		Pulse Rate & 0-150 BPM & 48 bps & 12 & 4 \\ \hline
		Respiratory Rate & 2-50  breaths/min & 240 bps  & 12 & 0.1-20\\ \hline
		Blood Pressure & 10-400 mm Hg & 1.2 Kbps& 12 & 0-100  \\ \hline
		Pathogen detection & 0-1 & 2.4-160 bps  & 12 &- \\ \hline
		Blood Temperature & 32-40 C & 2.4-120 bps& 12 & 0-1  \\ \hline
		Blood CRP & 0-8 mg/l & 2.4 bps& 12 &-   \\ \hline
	\end{tabular}
\end{table*}
\subsubsection{Bluetooth Low Energy}
	BLE has many lucrative features that can be important to on-body communications, such as low-power, low-rate and low-range \cite{51}. It operates in 2.4 GHz frequency band while the time needed for connection setup and data transfer is less than 3 ms. Furthermore, BLE offers a data rate of up to 1 Mbps which makes it a suitable choice for on-body communication.
	For example, in the HDT-based personalized elderly type 2 diabetes proposed by Thamotharan et al. in \cite{512}, BLE was used to transmitted vital signs, blood glucose levels, activity levels, and other information to the mobile phone. BLE, however, does not support multicast communication, which may be important for some HDT applications. 

\subsubsection{ZigBee}
	 It is developed atop the IEEE 802.15.4 standard for low-power, short-range and low-rate data connectivity \cite{75}. Unlike BLE, ZigBee supports various network topologies and a huge number of sensors, making it a more robust solution. Furthermore, ZigBee is known to be not only a secure key technology that offers three levels of security mode to prevent unauthorized access of data by attackers, but also capable of supporting multicast \cite{75}. However, ZigBee shares the frequency bands with other types of radio technologies (e.g., WiFi and Bluetooth), and therefore, suffers from unintentional interference. Additionally, ZigBee communications are vulnerable to radio jamming attacks due to the openness of the wireless medium. When a malicious device emits a high-power jamming signal, all ZigBee devices in its proximity will be unable to work \cite{68}. Therefore, its architecture requires a significant upgrade to be suitable for adoption in HDT.

\begin{figure}[!t]
	\centering
	\includegraphics[width=0.95\columnwidth]{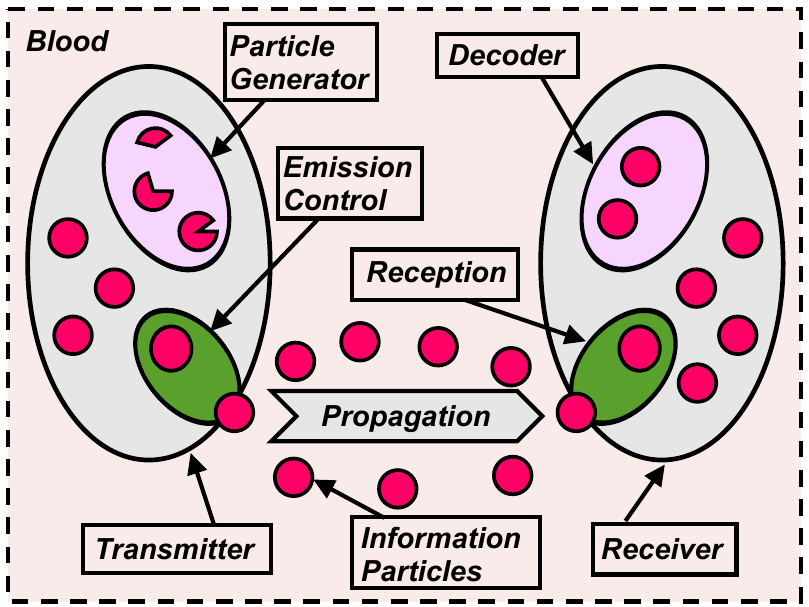} \\
	\caption{An illustration of the molecular communication.
	}\label{MC}
\end{figure}
Since conveying information using electrical or electromagnetic waves is impossible in small dimensions \cite{53}, radio technologies (e.g., BLE and ZigBee) cannot be used inside the human body, and this drives the emergence of molecular communication (MC). 
\subsubsection{Molecular Communication}
	MC is a bio-inspired communication method with the ability to mimic the communication mechanism of living cells \cite{54}. As shown in Fig. \ref{MC}, MC relies on the use of molecules for the transmissions and receptions of information. Specifically, a transmitter releases small particles, called information particles, which are typically a few nanometers to a few micrometres in size. An example is when releasing molecules or lipid vesicles into an aqueous medium (e.g., blood), tiny information particles propagate freely until such particles arrive at a receiver. Subsequently, the receiver detects and decodes the information encoded in these particles. Despite its numerous advantages, it is not clear how MC can establish interfaces for interconnecting human bodies and the external environment. Such interfaces are expected to possess the ability to convert chemical (or molecular) signals into equivalents (e.g., electrical and optical signals)  acceptable by conventional communication mechanisms \cite{77}. Besides this interface issue, multiple-input and multiple-output (MIMO) MC may be required to ensure real-time health parameters detection in HDT, while guaranteeing the protection of data security \cite{56}.

It is worth noting that, other wireless technologies, such as narrowband IoT (NB-IoT) and IPv6 over low-power wireless personal area networks (6LoWPAN) may also be alternatives for on-body communication in HDT. A review of these technologies can be found in \cite{295}.

\subsection{Beyond-Body Communication}\label{SE4_2}
HDT should be supported by the bidirectional real-time synchronization between any PT-VT pair to ensure high fidelity of VT \cite{518}. This synchronization is, however, data-driven and delay-sensitive. Furthermore, data captured through sensing devices in the physical world are often complex, massive, heterogeneous, multiscale, and with high noise. In addition to real-time synchronization, interacting with VTs involves more complex information that needs to be transmitted between the PT and users in the physical world. Multimodal information, such as 3D virtual items, text, images, haptic feedback, smells, among others, needs to be transmitted in HDT under various applications to enhance the immersive experience.
These specific characteristics place a significant burden on current communication networks. For instance, a massive number of PTs simultaneously interact with their respective VTs in the same area, which poses great challenges to bandwidth-limited communication to support the delivery of high-resolution contents \cite{481}. To address these issues, we provide a review of cutting-edge communication solutions that can enable beyond-body communication for HDT.

\subsubsection{Tactile Internet}
HDT-enabled healthcare applications require not only 360$^{\circ}$ visual and auditory content for immersive experiences but also haptic interactions \cite{511}. For example, in the virtual simulation of liver surgery for optimal surgery planning, visual and haptic feedback from the digital liver is essential for accurate evaluations of complex intrahepatic anatomical structures \cite{87}. However, these feedbacks require xURLLC service to ensure the timeliness necessary to facilitate efficient decision making in the physical environment. Delayed feedback may cause serious disruptions, such as deaths \cite{90}. Fortunately, tactile Internet (TI) is a promising solution, which facilitates fast multimodal interactions with multisensory information \cite{469}. Specifically, it can effectively transmit audio-visual-haptic feedback in real-time between real and virtual environments \cite{511}. This will ensure that PTs are immersed in virtual space in a holistic and multi-sensory way.
Beyond HDT, such a breakthrough in audio-visual-haptic feedback transmission will also change the way humans communicate worldwide.

Although TI has the potential to revolutionize the future of wireless communication, it is still far from being deployed on a large scale because of two major barriers. First, it is still difficult to establish a consensus on the performance of the TI, especially for large-scale implementations, owing to the lack of a TI testbed. Second, the overall progress of TI has been severely impeded due to the asynchronous efforts among different disciplines of TI. To address these critical issues, Gokhale et al. developed a common testbed, called tactile Internet eXtensible testbed (TIXT), for TI applications \cite{88}. To present the TIXT proof of concept, two realistic use cases belonging to two different classes -- human operator-machine teleoperator in a virtual environment and a physical environment -- were demonstrated. In \cite{89}, Polachan et al. designed and implemented a tactile cyber-physical systems (TCPS) testbed, called TCPSbed, to provide rapid prototyping and evaluation of TCPS applications. Different from the previous testbed without assessments, TCPSbed includes tools for the characterization of latency and control performance.

TI is expected to enable a paradigm shift from traditional content-oriented communication to control-oriented communication \cite{91}. However, such a paradigm has stringent requirements in terms of high reliability and sub-millisecond latency which pose daunting challenges for resource allocation in networks. Therefore, Gholipoor et al. investigated a joint radio resource allocation and networks function virtualization (NFV) resource allocation in a heterogeneous network \cite{92}. The authors jointly considered queuing delays, transmission delays as well as delays resulting from the execution of the virtual network function. Following this, the authors formulated a resource allocation problem to minimize the total cost function subject to guaranteeing end-to-end delay of each tactile user for this setup and proposed two heuristic algorithms to solve the problem. In addition, since cellular networks are resource constrained, accommodating haptic users along with existing non-haptic users becomes a hard scheduling problem. Therefore, Samanta et al. proposed an efficient latency-aware uplink resource allocation scheme to satisfy the end-to-end delay requirements of haptic users in a long-term evolution (LTE)-based cellular network \cite{93}.

In summary, TI is already paving the way for HDT, where PTs and VTs are connected via the extremely reliable and responsive networks of the TI to enable real-time interactions.

\subsubsection{Semantic Communication}
The explosive growth of data, expected in HDT, as well as the generally discussed limited bandwidth issues in wireless communications, indicates the necessity of revolutionizing the classical Shannon theory based solutions. Semantic communication is a possible solution for HDT applications and is being considered a disruptive technology with the ability to eliminate the limitations of the conventional data-oriented communication paradigm. Unlike traditions where channels with relatively infinite capacities are required to ensure real-time traffic, semantic-oriented communication-based paradigms allow information to be transmitted at the semantic level, rather than bit sequences \cite{79}. By integrating machine learning (ML) algorithms, knowledge representation and reasoning tools, semantic-oriented communication-based paradigms can facilitate semantic recognition, knowledge modelling and coordination \cite{80}. Generally, semantic communication extracts ``meanings'' of any transmitted information at the transmitter through the use of ML algorithms and encoding the extracted features with source knowledge base (KB). Then this semantic information is transmitted to the intending receiver and is successfully ``interpreted'' by the receiver using a matched knowledge base between such a transmitter-receiver pair and ML algorithms \cite{488, 459}.

\begin{figure*}[!t]
	\centering
	\includegraphics[width=0.935\textwidth]{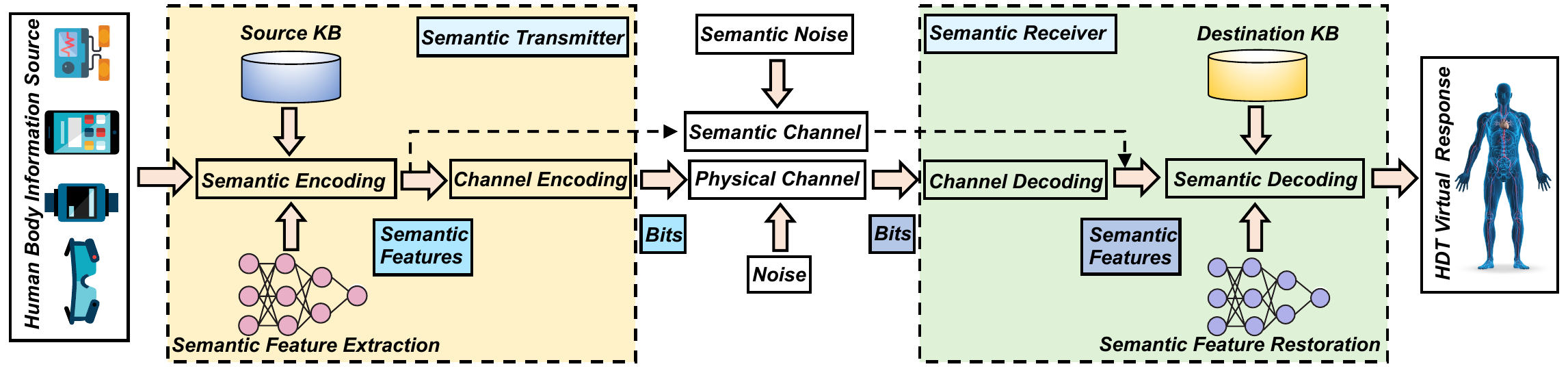} \\
	\caption{Semantic communication for HDT.}\label{seman}
\end{figure*}

As shown in Fig. \ref{seman}, semantic communication mainly consists of three components \cite{79} : 

\begin{enumerate}[-]
	\item \emph{Semantic transmitter (encoder)}: This component is responsible for the extraction and identification of the semantic features of each raw message. It also performs message compression as well as the removal of irrelevant information. After this, it encodes the obtained features into symbols (bits) for transmission.
	\item \emph{Semantic receiver (decoder)}: Semantic receiver decodes and infers semantic features in a format or structure that is understandable to the target user. 
	\item \emph{Semantic noise}:  Semantic noise usually interferes with semantic information during transmission and may result in misunderstanding or misperception of the semantic information at the intending receiver. It commonly appears in the procedures of semantic encoding, transmission and decoding.
\end{enumerate}

Semantic communication is recognized as a promising technology to support the wide proliferation of intelligent devices, such as XR devices and smartphones, in HDT with specific requirements of huge radio resources, time-sensitivity of transmitted data, low latency and high accuracy. 
We highlight three potential benefits that it can bring to HDT.
\begin{enumerate}[-]
	\item \emph{Alleviating the burden on data transmission in HDT:} Take the application of XR in HDT as an example. In the scenario of telemedicine through HDT, VTs of doctors and patients are presented in a digital environment through XR, which generate massive of data in various of forms (e.g., text, audio, images, video and haptic) to be transmitted. To guarantee the ideal immersive experience for users, the end-to-end latency and data rate requirements have to be strictly met. In the semantic communication paradigm, the data can be extracted semantically first. This allows XR devices to transmit the information concerned by the XR server for operation after understanding and filtering out the irrelevant information to save bandwidth and reduce computing latency at the XR server. Meanwhile, the XR server can also extract semantic information, ignoring irrelevant details in the face of bandwidth constraints, and thereby reducing downlink pressure. Moreover, with the decrease of the amount of actual bits transmitted, all intelligent devices in HDT can work in a more energy-efficient manner.
	Additionally, semantic communication can be used for efficient verification of PT-VT pair synchronization. The identification via channels is a recent and highly efficient semantic communication method that allows transmission of identifiers with doubly-exponential scaling in size with respect to the block length and code rate \cite{499,500}. It has been demonstrated that, via this approach, remarkable effectiveness may be achieved in verifying PT-VT pair synchronization for HDT \cite{501}.

	\item \emph{Promoting the data security and privacy in data transmission:} By pre-processing the source data in semantic communication, the communication parties (e.g., PT-VT pairs) only exchange semantic information extracted according to the communication tasks, instead of the complete source data, which can enhance the security of the network to a large extent. Moreover, communication parties in semantic communication are required to share their KBs to infer the semantic information. This hinders eavesdroppers to interpret valid information from the intercepted data without obtaining the specific KBs from the communication parties. It is of great significance to the privacy-sensitive health data transmitted in HDT. 
\end{enumerate}

The limited computation and storage capabilities of these intelligent devices, however, restrict the local implementation of complex and energy-intensive ML algorithms (e.g., deep neural networks (DNN) algorithms), training of semantic encoder, semantic decoder, channel encoder and channel decoder. One way to eliminate this issue is to simplify the structure of neural networks by performing model compression through the adoption of network sparsification and quantization. Xie et al. proposed a lightweight semantic communication system to support the transmission of low-complexity text in IoT \cite{82}. The presented approach removes redundant nodes and weights from the semantic communication model by adopting neural network pruning techniques, thus can reduce the required computational resources at IoT devices. With this, IoT devices can take advantage of semantic communication, thereby lowering the required bandwidth when transmitting to the cloud/edge. Besides, federated learning (FL) and distributed learning are other learning-based techniques that can facilitate efficient training of any ML-based semantic communication system. The authors in \cite{83} proposed an FL-enabled scheme to support the information semantics of audio signals. The presented FL-enabled solution allows a joint training of federated semantic communication models among IoT devices without sharing sensitive information.

However, the lack of appropriate performance metrics is a critical issue in semantic communication systems. Unlike the traditional communication methods which often focus on minimizing bit-error-rate (BER) and symbol-error rate (SER) to ensure that more bits can be transmitted with fewer communication resources, semantic communications are more complicated. The performance metrics of semantic communication are diverse and depend on the type of messages. In \cite{84}, sentence similarity was proposed as a suitable metric to measure the semantic error of any transmitted sentence. Similarly, a peak signal-to-noise ratio (PSNR) was used to evaluate the performance of an image semantic communication system \cite{85}. 
Indeed, the design guidance for semantic communication is still provided by the Shannon theory. With such a guide, semantic information can be encoded into bit streams, and then transformed into physical signals before being transmitted via communication channels. As a result, various existing traditional signal processing schemes can support semantic communications, while advanced wireless communication technologies are expected to enable more efficient semantic communication systems \cite{79}. 

\subsection{Lessons Learned}\label{SE4_3}
Communication in HDT involves both on-body and beyond-body communications. On-body communication utilizes wireless technologies such as BLE, ZigBee, and MC to transmit data around the human body, enabled by WBAN. Beyond-body communication connects the PT and its corresponding VT, utilizing cutting-edge technologies such as TI and semantic communication to transmit the massive multimodal data.

While these communication technologies can enable the communication layer in HDT, there are several opening issues remaining. First, since multiple wireless communication technologies may be implemented simultaneously, interference among them should be addressed to ensure reliable communications. Second, efficient resource optimization of communication resources is crucial to enable effective interaction between PT and VT. Resource optimization should aim to balance the use of available resources and minimize the delay in data transmission.
Third, ensuring the security and privacy of healthcare-related data during transmission is critical. This involves implementing appropriate encryption mechanisms and secure communication protocols to prevent unauthorized access to sensitive patient data. Additionally, different measures should be taken to ensure the integrity of the data, such as digital signatures and secure timestamps.

In summary, addressing these communication-related issues is essential to ensure the successful implementation of HDT. Future research should focus on developing innovative solutions for interference management, resource optimization, and security and privacy protection to enable reliable and secure communication in HDT.

\section{Key Technologies for Computation Layer} \label{SE5}
Building HDT requires the assistance of powerful computing systems. For example, achieving a profoundly immersive experience through real-time rendering of PT, enabling interaction-driven optimization through fast data analysis (e.g., the responsive facial expression during interaction \cite{514}), all of these require the support of unparalleled computing power. However, the current central architecture based computation paradigm cannot meet the fast-responsive and computation-intensive requirements of HDT. Edge computing is capable of providing real-time and supplementary computing capability, and thus attracts a  lot of attentions in HDT applications.

HDT generally relies on extensively complex AI algorithms to continuously update each VT for enhancing reliable diagnoses, predictions and accurate decisions to support counterpart PT in the physical environment. To achieve this, HDT requires accurate massive data and intensive computation power, which can hardly be met by resource-constrained mobile and IoT devices. One may consider to enable task offloading that allows high computation-demanding tasks to be offloaded to more powerful cloud computing systems such as  AWS, AliCloud and Azure, to mitigate the computation pressure on the mobile devices. 

\begin{figure*}[!t]
	\centering
	\includegraphics[width=0.95\textwidth]{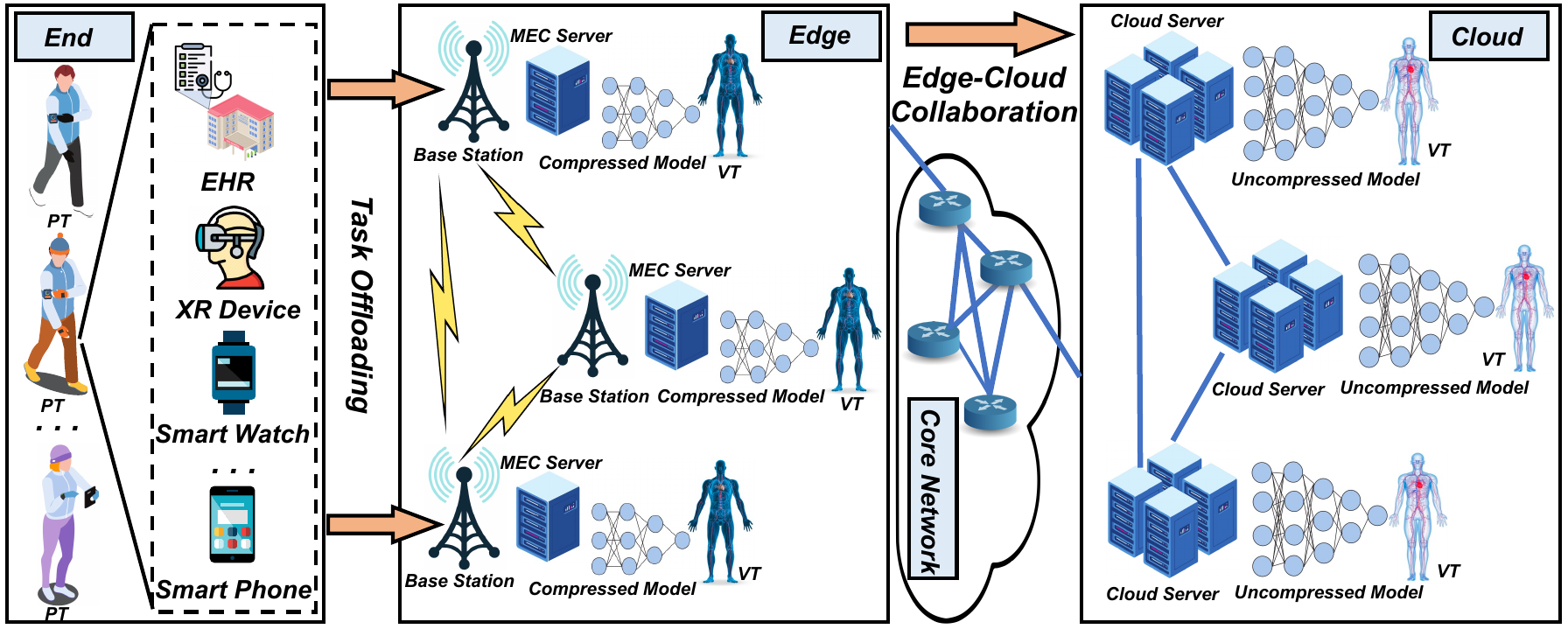} \\
	\caption{The edge computing paradigm for HDT.}\label{MEC}
\end{figure*}

However, cloud computing suffers from many limitations: i) high communication cost due to the high distance between cloud servers and users; ii) network congestion due to the massive amount of data that are simultaneously transmitted over the network; iii) user data security and privacy issue due to centralized storage. 
These motivate the continuous emergence of edge computing as a promising, practical and efficient solution because of its ability to bring computing capabilities near users, thereby alleviating the need to transmit data to the central cloud for computation. 
Fig. \ref{MEC} presents the edge computing paradigm for HDT.

In this section, we firstly discuss the role of mobile edge computing in HDT, and its benefits, i.e., addressing the mobility issue of HDT and reducing the response time, in Section \ref{SE5_1}. Although mobile edge computing can be an ideal solution for replacing the traditional centric cloud computing, it hard to resolve computation burden brought by long-term operations and large scalability of HDT under several scenarios. In response, we further introduce an effective computing paradigm, called edge-cloud collaboration, in Section \ref{EC-CLO}. Finally, we summarize the reviewed papers in Section \ref{SE5_3}, and discuss several opening issues therein.

\subsection{Multi-Access Edge Computing}\label{SE5_1}
Mobile edge computing (MEC) extends cloud computing capabilities to the edges of the ubiquitous radio access networks (e.g., 3G/4G macro-cell or small-cell base stations) that are close to mobile users \cite{100}. With the capability of providing pervasive, prompt and agile computation services anytime and anywhere \cite{102, 480}, MEC continues to promote its adoption in HDT. 
\subsubsection{Addressing the Mobility Issue of HDT}
With the adoption of MEC, the mobility of HDT can be well addressed. Martinez et al. presented a cardio twin architecture for the detection of ischemic heart disease (IHD) \cite{95}. Since the IHD is time sensitive, an edge computing paradigm was integrated into the cardio twin design. The design was implemented on a modern smartphone with sufficient computing capabilities. This makes the HDT implementation highly responsive and accessible with the ability to cope with mobility. The proposed HDT design collects data from multiple sensors, medical records and social networks while performing DL classification models for detection, prevention and reduction of IHD risk. In a similar work, Díaz et al. in \cite{103} proposed a personalized HDT coach for physical activities to improve training performance. The proposed HDT was implemented on an edge platform. 
By optimizing ML algorithms, the HDT coaching system was able to derive the performance of a trainee (i.e., the similarity score between the coach and the trainee) in real-time based on the estimation of trainee's pose, obtained through a camera integrated with the edge device. 

\subsubsection{Reducing the Response Time}
By adopting MEC, the response time of HDT can be significantly reduced, particularly the latency of feedback from the VT can be decreased compared to using cloud computing alone \cite{490}. This was verified by the authors in \cite{124}, where two case studies were carried out. In this work, a novel edge-based architecture for PH monitoring, named BodyEdge, was proposed, consisting of two complementary components, i.e., a software client installed on mobile devices and a hardware edge gateway located at the edge of the network. The software client acts as a multi-radio communication relay node and enables the body sensor network (BSN) to reach the edge gateway. The hardware edge gateway can be deployed on resource-constrained hardware platforms supporting multi-radio and multi-technology communication. The performance of the proposed architecture was evaluated based on its ability to detect cardiac for users in two different scenarios, i.e., workers operating in a factory and athletes training in a fitness center. As expected, the obtained results demonstrated that the response time of the  centralized computing architecture is more than doubled compared to that of BodyEdge.

\begin{table*}[!ht] 
	\centering
	\caption{QoS Requirements for HDT Applications (Compiled from \cite{463, 465})}
	\label{qos}
	\begin{tabular}{|m{3cm} <{\centering} | m{3cm} <{\centering} | m{3cm} <{\centering}|m{3cm} <{\centering} | m{3cm} <{\centering} | m{3cm} <{\centering} |}
		\hline
		\rowcolor{lightgray}\textbf{Use Case} & \textbf{Link Data Rate} & \textbf{Latency} & \textbf{Jitter}  & \textbf{Reliability} \\ \hline
		Health Monitoring & $\geq$ 1 Gbps & $\leq$ 500 $\upmu$s & $\leq$ 10 $\upmu$s &  $\geq$ 99.9999 $\%$\\ \hline
		Diagnosis & $\geq$ 10 Gbps & $\leq$ 1 ms & $\leq$ 100 $\upmu$s  & $\geq$ 99.99999 $\%$ \\ \hline
		Surgery & $\geq$ 100 Gbps & $\leq$ 500 $\upmu$s & $\leq$ 10 $\upmu$s & $\geq$ 99.999999 $\%$ \\ \hline
		Rehabilitation & $\geq$ 10 Gbps & $\leq$ 5 ms & $\leq$ 100 $\upmu$s & $\geq$ 99.999 $\%$  \\ \hline
	\end{tabular}
\end{table*}

\subsubsection{Achieving the Extremely High Quality-of-Service (QoS)}
With the adoption of MEC, the extremely high QoS of HDT can be achieved. As presented in Table \ref{qos}, the QoS requirements for HDT applications are relatively stringent, necessitating the assistance of MEC. For example, AR is a critical and widely applicable technology for HDT in monitoring, diagnosis, surgery and rehabilitation, with the ability to enable an immerse interaction of users with VTs. However, AR applications are considerably computation-intensive, putting a substantial computational burden on mobile devices. Therefore, offloading AR applications to edge nodes can alleviate the computational burden and reduce the latency of AR services, thereby improving the experience of users in HDT. Peng et al. in \cite{104} explored the possibility of offloading the computation of AR applications in HDT to edge nodes while considering users' privacy protection and mobility. A novel multi-objective meta-heuristic method was proposed, aiming to preserve privacy, minimize the motion-to-photon latency and energy computation while maintaining load balancing among edge nodes.

However, several nontrivial challenges also arise for practical implementation of MEC in HDT. These include high energy consumption, straining radio resources, high computation burden on edge servers and data privacy. To this end, some research efforts have been dedicated \cite{111, 112, 113, 114, 457, 460}. For example, an MEC-enabled framework with the ability to support multi-user computation offloading and transmission scheduling for delay-sensitive applications was proposed in \cite{111}. By considering tradeoffs between local and edge computing, this work proposed a novel mechanism to jointly determine the computation offloading scheme, transmission scheduling discipline and pricing rule. The results showed that the proposed mechanism can ensure that no mobile user has the incentive to strategically deviate the network-wide optimal management and maximize the network social welfare. Furthermore, the authors in \cite{112} investigated the workload re-allocation for edge computing with server collaboration.
To achieve the equilibrium for each edge server via minimizing expected costs (including energy consumption, delay, transmission, configuration and pricing costs), a novel cooperative queueing game approach was proposed. 
Bishoyi et al. in \cite{114} focused on joint cost and energy-efficient task offloading in the MEC-enabled healthcare system by designing optimal incentive schemes for HDT to curtail the amount of task offloading. Particularly, the interaction among MEC servers and HDT users was modelled using the Stackelberg game to derive the optimal task offloading decision for HDT users while activating the corresponding reimbursement amount based on the amount of local computing task.

Although MEC is a promising alternative to traditional centralized cloud computing, the limited resources available on edge servers (such as computation and storage resources) make it difficult to handle long-term, large-scale HDT data and extremely massive and intensive computation tasks. For example, implementing an HDT brain requires 10,000 graphical processing unit (GPU) cards (AMD, 16 GB Memory) to simulate 86 billion neurons \cite{523}. Fortunately, the edge-cloud collaboration paradigm has recently emerged.

\subsection{Edge-Cloud Collaboration} \label{EC-CLO} 
Edge-cloud collaboration combines the benefits of both edge computing and cloud comping, thereby providing a promising solution towards addressing the limitations of enabling either the edge or cloud along \cite{117}. In the edge-cloud computing architecture, an end-user can offload complicated tasks to an edge server. Each edge server can then decide whether to compute the whole task or collaborate with the cloud server to perform a part of it. The edge-cloud collaboration paradigm can reduce overall latency and augment computation capacity \cite{118, 498}. This advantage is very useful in HDT for optimizing the information processing procedure. 
 Ghosh et al. in \cite{125} proposed an IoT-edge-fog-cloud-based framework to provide PH services to users with minimum delay. The system used IoMT devices in BSN to capture the medical data of each PT. A smartphone was used as the edge device to accumulate geo-location information of users, health data and contextual data including the humidity, light intensity and temperature. The accumulated data was initially processed inside a fog device (e.g., roadside unit (outdoor) or small cell cloud enhanced eNodeB (indoor)) before being forwarded to the cloud. The cloud server then carried out an analysis to detect any abnormality and subsequently send an appropriate alert to users.

Edge-cloud collaboration ensures data distribution across multiple edges, thereby reducing the risk of privacy leakages.

\begin{figure}[!t]
	\centering
	\includegraphics[width=0.95\columnwidth]{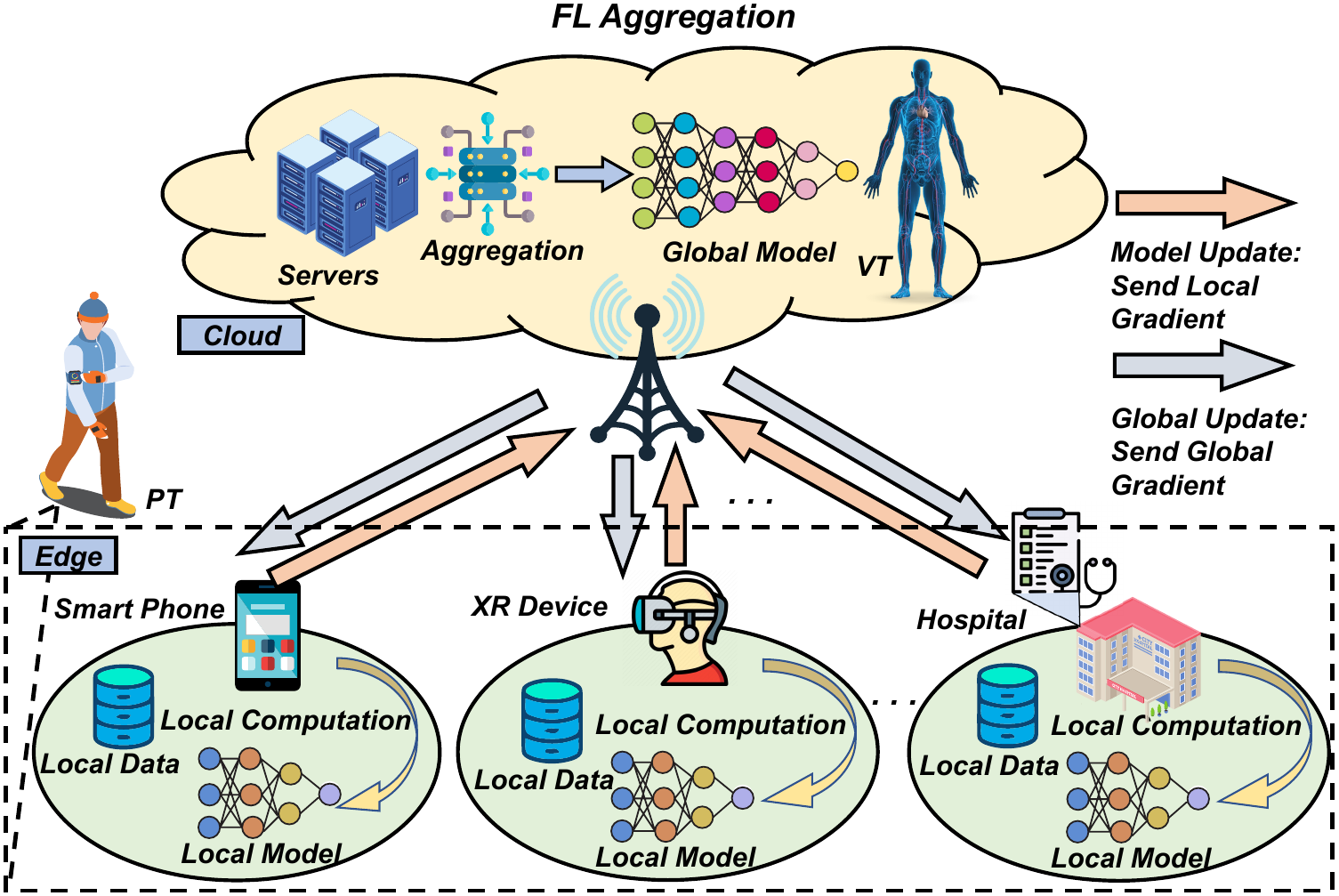} \\
	\caption{Edge-cloud collaboration assisted FL.}\label{FL}
\end{figure}

FL is recognized as a good match that can be adopted in edge-cloud computing platforms to further improve the data privacy. Generally speaking, FL is a distributed learning paradigm that allows each local device to perform local training on its data while sharing the model updates rather than raw data with the FL server for aggregation and update of the global model \cite{485,486}. In the edge-cloud collaboration assisted FL framework, local model training can be achieved at the edge, while model aggregation is carried out at the cloud, as shown in Fig. \ref{FL}. 

This edge-cloud collaboration assisted FL framework can significantly improve data security and privacy of HDT. For instance, Okegbile et al. in \cite{527} proposed a differentially private federated multi-task learning framework for HDT that integrates three key techniques – differential privacy, federated multi-task learning, and blockchain – to enhance the security, privacy, and efficiency of human-to-digital twin connectivity. Besides, anomaly detection (AD) in a centralized healthcare ecosystem is often inherently plagued by privacy and security issues (e.g., data poisoning) when sending medical data of patients to a centralized server. To address this issue, Gupta et al. in \cite{110} proposed an FL-based AD model by utilizing edge cloudlets to locally execute AD models without sharing the data of each patient. A novel hierarchical FL was introduced to allow aggregation at different levels following multi-party collaboration (e.g., disease-based grouping or age-based grouping). This multi-party collaboration eliminates the limitations of the traditional single aggregation server. Additionally, the authors presented a proof-of-concept implementation for HDT, where the data from each patient is forwarded to the AD model, located in the edge cloudlet, to detect any anomaly. Similarly, Wu et al. in \cite{123} proposed a novel edge-cloud-based FL framework for personalized in-home health monitoring. From multiple homes at the network edges, the proposed framework can learn a shared global model in the cloud and ensures data privacy through the adoption of FL. 

\subsection{Lessons Learned}\label{SE5_3}
The implementation of HDT will result in a series of computationally intensive tasks that require a powerful and fast-responsive computing service. In order to address the challenges associated with HDT, researchers are exploring the potential of using MEC and edge-cloud collaboration as promising paradigms.

However, there are still several opening issues that need to be considered. First, HDT may requires massive computing resources for an individual task (e.g., evolution of PT), let alone under a worldwide scenario, which significantly exacerbates the computing burden on the network side. Therefore, it is imperative to develop novel computing resource allocation schemes for MEC-based or edge-cloud collaboration-based HDT. These schemes should aim to balance the computing loads across different computing resources and optimize the utilization of available resources.
Second, in the context of distributed learning, such as FL, effective incentive mechanisms are needed to attract more servers to participate in the computing tasks of HDTs. These incentive mechanisms can be designed to compensate servers for their contributions to the computing tasks, such as offering high payments or recognitions for their participation. In addition, the design of these incentive mechanisms should consider the potential risks associated with the participation of servers, such as privacy leakages, and reduce these risks through appropriate privacy-preserving methods.

\section{Key Technologies for Data Management Layer} \label{SE6}
Since every segment in HDT generates and exchanges massive data constantly, data management for HDT is indispensable. For guaranteeing the efficiency of such huge data stream, it is required that data should be pre-processed, and then stored in databases for later analysis. On top of these, data security and privacy schemes, such as cybersecurity, privacy-preserving mechannism and distributed ledger technology, are also imperative in the data management for HDT. 

\begin{figure}[!t]
	\centering
	\includegraphics[width=0.99\columnwidth]{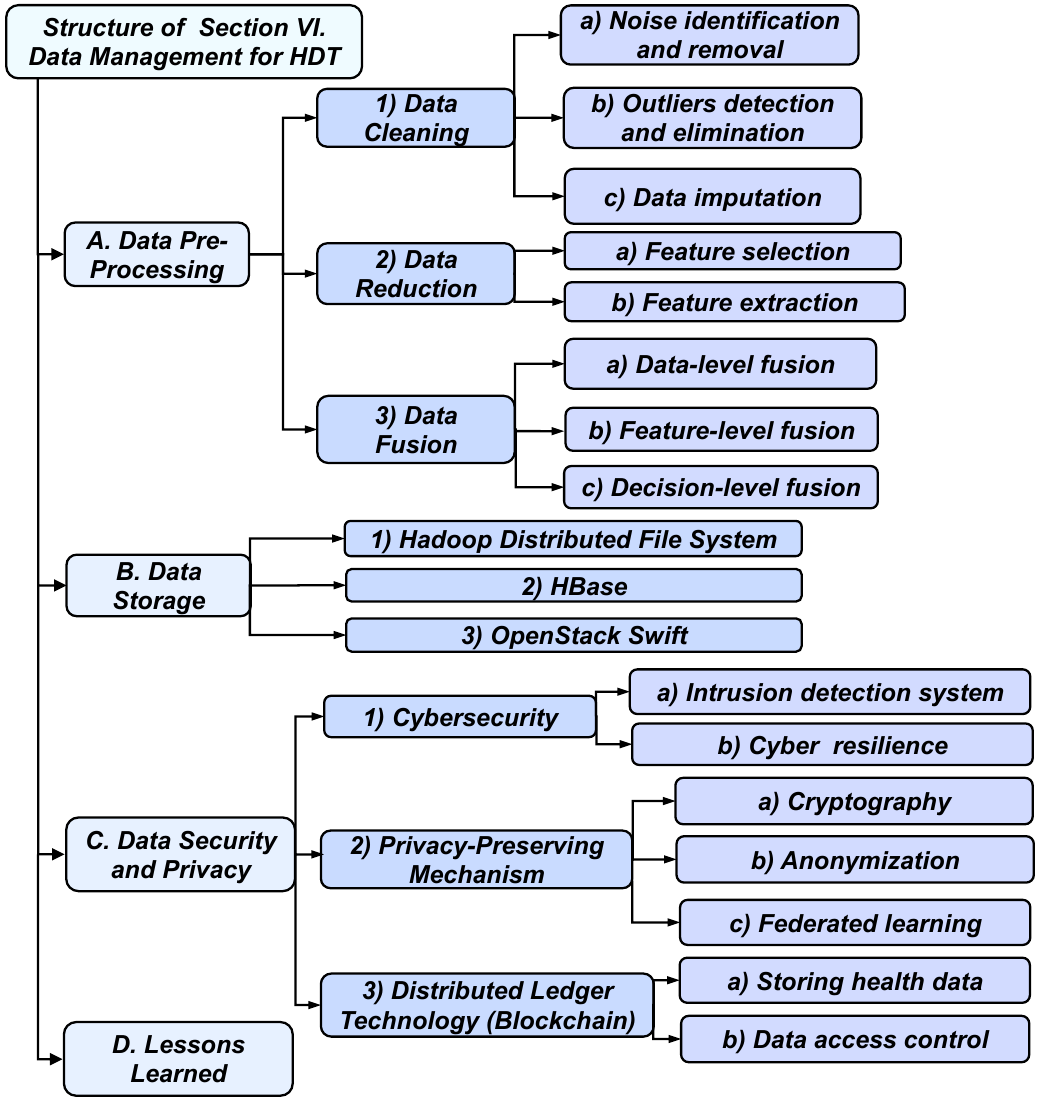} \\
	\caption{The roadmap of Section \ref{SE6}.}\label{Stru_6}
\end{figure}

In this section, we survey data management for HDT from three perspectives, namely, data pre-processing, data storage, and data security and privacy. Specifically, we begin by discussing data pre-processing in Section \ref{SE6_1}, which includes data cleaning, data reduction, and data fusion. Then, we present data storage in Section \ref{SE6_2}. In Section \ref{SE6_3}, we discuss data security and privacy for HDT, which includes cybersecurity, privacy-preserving mechanisms, and distributed ledger technology. Finally, we provide a brief summary of this section, followed by a discussion of several opening issues in Section \ref{SE6_4}. We provide a roadmap of this section, as shown in Fig. \ref{Stru_6}.

\subsection{Data Pre-Processing}\label{SE6_1}
As previously mentioned, physical data in HDT have the characteristics of heterogeneity, multiscale and high noises. Hence, it is necessary to enable pre-processing such that issues like missing data, data redundancy, data conflicts and errors can be properly handled \cite{387}. Generally, data pre-processing in HDT includes data cleaning, data reduction and data fusion.

\subsubsection{Data Cleaning}
Data cleaning mainly consists of noise identification, noise removal, outliers detection and elimination, and data imputation. 
	\begin{enumerate} [a)]
	\item \emph{Noise identification and removal}: Data collected in HDT are commonly contaminated by noise, which may degrade the data quality, thereby affecting the overall performance of HDT. Chiang et al. in \cite{407} proposed a fully convolutional network (FCN)-based denoising autoencoder (DAE) to reconstruct the clean ECG signals from its noisy version. Xu et al. in \cite{408} adopted median filtering to facilitate the removal of noise in medical images. Nasrin et al. in \cite{409} applied a recurrent residual U-Net (R2U-Net) based autoencoder model to denoise medical images.
	
	\item \emph{Outliers detection and elimination}: An outlier is any data object that deviates significantly from the rest and thus its removal can improve the overall performance of HDT. K-means clustering algorithm can be adopted to detect outliers in health data \cite{410}. Another popular outliers detection and elimination technique for health data are density-enabled spatial clustering of applications with noise (DBSCAN)-based method, which has been used to detect and eliminate outliers in heart disease datasets \cite{411}. 
	
	\item \emph{Data imputation}: In HDT, massive medical data are collected from multiple sources to achieve PT-VT synchronization. Hence, such a system sometimes suffers from missing data. 
	To deal with this, the data imputation technique – a technique that replaces all missing data based on the structure of the underlying datasets -- has been demonstrated to be an effective solution \cite{130}.
	
	\begin{itemize}	
	\item K-nearest neighbors imputation (KNN-imputation) algorithm is an effective technique when applied in the field of bioinformatics to impute missing data.

	For example, Barricelli et al. in \cite{130} proposed a team of HDTs, where each HDT tracks fitness-related measurements that describes the behaviour of an athlete on consecutive days. Since missing data are inevitable in such a system, the authors adopted the KNN-imputation technique to impute the missing data for guaranteeing robustness to noisy data while maximizing the data informativeness. A similar method was adopted in \cite{134}, where Zhang et al. employed the KNN-imputation technique to replace the missing data in an HDT, to support the diagnosis of lung cancer.

	\item DL-based imputation (DLI) technique has shown its superiority when filling missing data, especially for discrete ones. KNN-imputation method relies on data continuity, while the medical data in HDT are sometimes discrete. DLI can complete missing values at the character level owing to its strong nonlinear approximation capabilities \cite{405}. Hence, such a technique can provide better precision compared with the traditional imputation methods and is suitable in HDT with discrete data \cite{316}. 
	Tai et al. in \cite{316} adopted a DLI filling-enabled technique to limit the impact of missing health data in HDT-based surgery simulation. Phung et al. in \cite{406} proposed a DLI technique, called the overcomplete denoising autoencoder (ODAE) model, to address the missing data problem in health datasets. The proposed ODAE model consists of two components, i.e., a deep denoising autoencoder that extracts the correlations between variables and a modified loss function, which prioritizes learning the “real” distribution of each variable.
	
\end{itemize}
	
%
\end{enumerate}

\subsubsection{Data Reduction}
Since data collections in HDT are usually from multiple sources, such data are expected to include invalid, non-informative and redundant data. It is thus essential to adopt appropriate data reduction techniques to reduce the data dimensionality while maintaining the integrity of the data and keeping the most informative data. Data reduction is commonly achieved by feature selection and feature extraction.

\begin{enumerate}[a)]
	\item \emph{Feature selection}: Feature selection aims to select a subset of relevant features while removing irrelevant and redundant ones, to describe an object accurately without inducing much loss of information. The existing feature selection methods can be classified into three categories, namely filters, wrappers and embedded methods \cite{423}.	
	
	\begin{itemize}

	\item Filters evaluate the relevance of attributes without involving any learning algorithms (induction algorithms). As a result, filters are not computationally costly and possess a good generalization capacity. 
The widely discussed advantages of filters continue to motivate their usage by many researchers to facilitate the feature selection process in HDT. For instance, Alirezanejad et al. in \cite{426} proposed two heuristic filter methods -- Xvariance and mutual congestion -- for gene selection in medical datasets. Yuan et al. in \cite{427} proposed a new multivariable-synergy filter-based feature (gene) selection method, called partial maximum correlation information (PMCI), for microarray data. With the PMCI, the authors defined a feature importance indicator – variable importance in projection angle (VIPA) -- which was then used for ranking features when selecting the feature subset.

\item Wrappers require a performance evaluation of any data analysis model to determine the relevance of a selected feature subset. While interactions with data analysis models (learning models) make wrappers more computationally costly than filters, wrappers tend to perform better \cite{424}. Many studies have devoted in  wrapper-based approaches for feature selection of HDT data. For instance, Almugren et al. in \cite{429} proposed a wrapper-based feature selection algorithm for classifying cancer microarray gene expression profiles. The algorithm uses the FireFly algorithm along with a support vector machine (SVM) classifier named FF-SVM. In an early diabetes prediction task \cite{430}, Le et al. proposed a novel wrapper-based feature selection method using grey wolf optimization (GWO) and an adaptive particle swarm optimization to optimize the multilayer perceptron (MLP) with the aim of reducing the required input features. 

\item Feature selection in embedded method is performed as part of the model construction process.  
By this feature selection method, the search for the best subset of features is performed during the training process. Some efforts have adopted embedded methods to facilitate the feature selection process in HDT. For instance, Kang et al. in \cite{432} modified Lasso (a classic embedded method) for tumour classification in multi-class datasets. In \cite{434}, Li et al. proposed an improved SVM-RFE (another classic embedded method), called RFE with variable step size feature selection for microarray data, where the step size decreased with the number of selected features, so that the time consumption can be greatly reduced.
\end{itemize}
	\item \emph{Feature extraction}: Feature extraction is also known as feature projection, aiming to extracts a set of new features from the original ones. 
	\begin{itemize}

	\item Principle component analysis (PCA) is an unsupervised linear transformation technique commonly used for feature extraction and dimensionality reduction. Recently, several studies have used PCA as a feature extraction technique for classification in PH. For instance, R.M. et al. in \cite{442} proposed a Grey-Wolf optimizer by combing the PCA and the bio-inspired algorithm to extract the high-impact features from HDT datasets. Reddy et al. in \cite{443} employed PCA to extract the most important features from the cardiotocography dataset before inputting such datasets into ML algorithms for disease prediction.

	\item  Convolutional neural network (CNN) has become the most popular feature extraction technique and is an efficient method with the ability to reduce data dimension and produce a less redundant data set. 
To leverage the efficiency of CNN, Hurr et al. in \cite{438} adopted the CNN algorithm to aid feature extraction of ECG signals. Liu et al. in \cite{440} proposed a CNN-based feature extraction solution to support medical data feature advancement. Yang et al. in \cite{437} carried out feature extractions of tumour images by using two CNN models, namely Xception and DenseNe.

 \end{itemize}
	 

\end{enumerate}

\subsubsection{Data Fusion}

\begin{figure*}[!t]
	\centering
	\includegraphics[width=0.95\textwidth]{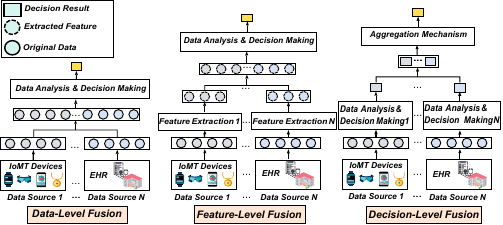} \\
	\caption{An illustration of different data fusion techniques for HDT.}\label{DF}
\end{figure*}

Data fusion refers to the process of merging data from several heterogeneous sources. Since in HDT, data are collected via heterogeneous sources, the data fusing process is a necessity. These multisource data may have unequal distribution of data classes, where one class has more instances than the others. Moreover, any data analysis that mines useful information from massive data needs more comprehensive data rather than one type of data. Hence, data fusion can enhance data collection, transmission, correlation and synthesis of useful information from various information sources in HDT.
Conceptually, the data fusion level in HDT can be categorized as: \emph{data-level fusion}, \emph{feature-level fusion}, \emph{decision-level fusion} \cite{420}, as summarized in Fig. \ref{DF}.


\begin{enumerate}[a)]
	\item \emph{Data-level fusion}: Data-level fusion is a low-level fusion that combines the raw data of users from multiple sources. 
	Such a low-level fusion is considered the simplest method to achieve a combination of inputs and can be performed at edge devices \cite{421}. Generally, the data in this fusion level are rearranged into a new matrix, where all the data collected from heterogeneous sources are placed next to each other. Recently, data-level fusion has been widely adopted in HDT frameworks. For example, Thung et al. in \cite{447} carried out an image-image fusion of PET and MRI images. The approach concatenated clinical and imaging features into one single feature vector before feeding them into the neural networks. However, data-level fusion usually contains a large amount of redundant data, and thus is not effective when adopted alone.

	
	\item \emph{Feature-level fusion}: Feature-level fusion is generally known as middle-level fusion and can be used to combine features obtained from multiple sources. The features are extracted by a preliminary feature extraction scheme to maintain relevant data while eliminating the insufficiently diverse and non-informative data from the datasets \cite{422}. It is, therefore, unsurprising that this paradigm has received a lot attentions in HDT applications. For instance, Ali et al. in \cite{431} proposed a novel framework that integrated both feature extraction technique and feature-level fusion to support heart disease prediction. In the proposed framework, the data of each heart patient were collected through various IoMT devices. After that, valuable Framingham risk factors (FRFs) were extracted from the unstructured data. Next, the proposed feature-level fusion approach was used to accurately combine FRFs and data collected via IoMT devices to generate more complete healthcare data for heart disease. An information gain (IG) approach was also developed to facilitate feature selection, while adopting a conditional probability tool to compute specific weights for heart disease features to increase the prediction accuracy. Finally, an ensemble DL classier was trained based on the fused features to predict heart disease in patients.
	
	\item \emph{Decision-level fusion}: Decision-level fusion generally refers to the process of leveraging predictions (decisions) from multiple models to make a final decision. Typically, different modalities are used to train separate models and the outcomes are combined into a complex model to aid final decisions. Thus, decision-level fusion can provide a global view and has also been extensively studied in HDT applications. To name a few, Yoo et al. in \cite{450}, through the mean value of predicted probabilities from two single modality models, obtained the final decision for the identification of patients with early multiple sclerosis (MS) symptoms. Qiu et al. in \cite{451} trained three independent DL-MRI models and applied voting techniques to combine them, thereby generating a fused DL-MRI model for mild cognitive impairment diagnosis.
		
%

\end{enumerate}

\subsection{Data Storage}\label{SE6_2}
HDT is expected to produce massive amounts of data including the demographics of PTs, clinical and medication history data, real-time personal health data obtained through pervasive sensing, genetic data, diagnostic trajectory, data analysis results and social and geographical relocation data. This requires massive storage to support various operations in HDT \cite{387}. Fortunately, the development of big data storage frameworks such as the MySQL database, HBase and NoSQL database has opened many opportunities to overcome this challenge in HDT.  

	\subsubsection{Hadoop Distributed File System (HDFS)}Hadoop is considered as the most famous platform for big data analytics. It uses a prime data storage system, called HDFS, to store massive data in distributed environments \cite{393}. HDFS has two data categories: metadata and application data. These data categories are stored separately. To be specific, the metadata are stored on a dedicated server, called the NameNode, while the application data are stored on other servers called DataNode \cite{394}. All these servers are mutually connected and communicate using TCP-based protocols. Owing to the benefit in improving data storage reliability, space utilization and fault tolerance, HDFS has been adopted to handle the storage of large-scale and complex HDT data. Babu et al. in \cite{395} employed HDFS to store structured, unstructured and semi-structured data from various sources. Gupta et al. in \cite{396} designed a Hadoop image processing interface (HIPI) to process the massive amount of medical imaging data simultaneously over a distributed cluster environment.

\subsubsection{HBase} HBase is an open-source, non-relational distributed database model with the ability to provide row-level queries and real-time read/write access to data \cite{393}. Each table in HBase is stored as a multidimensional sparse map, with rows and columns, where each cell has a time stamp. Hence, a cell value is uniquely identified. HBase offers both linear and modular scalability. Besides, it strictly maintains consistency of read/write operations which in return assists in automatic failover. These make HBase a popular tool for horizontal scaling of huge datasets \cite{393} and is a possible solution for HDT to enhance data storage. Addakiri et al. in \cite{397} proposed an HBase and MapReduce-enabled distributed healthcare data storage method, which was able to write data in a batch insertion manner, thereby achieving a better performance compared to traditional single insertion methods.

\subsubsection{OpenStack Swift} OpenStack Swift, also known as OpenStack object storage, is an open source object storage system, which provides a distributed, eventually consistent virtual object store for OpenStack. Swift can store billions of objects distributed across nodes and is capable of archiving and streaming \cite{400}. Besides, it is significantly scalable in terms of both size (amounts of bytes) and capacity (amounts of objects), and thus has been adopted in HDT to store regular healthcare data. Akintoye et al. in \cite{401} proposed a multi-phase data security and availability (MDSA) protocol to secure and improve the availability of healthcare data stored in the cloud. The authors implemented the proposed MDSA protocol in OpenStack architecture, where the healthcare data stored in Swift were secured and recoverable in case of a possible occurrence of an adversary attack or cloud server failure.


\subsection{Data Security and Privacy}\label{SE6_3}
HDT environments may suffer from various security threats.
Data security and privacy are important aspects of HDT implementations that require careful considerations \cite{507, 508}. Next, we discuss the main requirements of HDT from data security and privacy perspectives. 
\begin{enumerate}[-] 
	\item \emph{Confidentiality:} It implies that both obtained health data and information, transmitted among PTs and VTs, are only accessible by authorized end-users, and is usually achieved through the use of encryption and decryption algorithms.
	\item \emph{Data access control:} It defines a privacy policy intending to prevent unauthorized access to user information. Data access mechanisms are often set up with different access rights.
	\item \emph{Data integrity: } It requires that data accuracy (i.e., correctness and consistency) throughout the whole transmission process (e.g, data sharing) cannot be changed by unauthorized users.
	\item \emph{Data authentication:} It ensures proper authentications of participating users. In the absence of a reliable data authentication scheme, a malicious user may appear to be legitimate, thereby exchanging false and misleading data in the system. 
\end{enumerate}

Hereafter, we particularly survey some key technologies for protecting security and privacy of HDT.

\subsubsection{Cybersecurity}
It has been revealed that cyber-attacks are the most frequent causes of medical data breaches. An adversary can violate HDT privacy and security requirements by introducing ransomware attacks, stealing users’ information and blackmailing patients. For example, attackers may launch attacks on wireless personal area networks (WPANs), such as replay attacks on BLE, and eavesdropping on ZigBee \cite{505, 504}, to compromise the HDT resulting in organ hijacking. Some typical cyber-attacks in the context of HDT are elaborated as follows.

\begin{enumerate}[-]
	\item \emph{Denial of service (DoS)}: Attackers can flood WPANs with excessive traffics or requests, resulting in DoS attacks that can overload or disrupt the network's operations \cite{479}. This can cause devices such as wearable biomedical devices and smartphones in the data acquisition layer of HDT to disconnect, and the affected users may become unable to receive or send information.
	\item \emph{Battery exhaustion attacks}: WPAN devices, including wearable and implantable biomedical devices, are usually battery operated devices that enter sleep mode when inactive. Battery exhaustion attacks force continuous fraudulent connection requests to drain the battery and cause the device to become unavailable \cite{479}.
	\item \emph{Man in the middle (MITM) attacks}: An attacker may insert itself into the communication (e.g., BLE and ZigBee) channel between two legitimate devices (e.g., between personal devices or devices and gateways), while maintaining the facade that they are communicating with other directly \cite{479}. For instance, both legitimate generic access profile central and peripheral devices will be associated with the impostor device to monitor the messages between the two legitimate devices. 
	\item \emph{Data modification}: In HDT, an attacker may launch data modification attack, 
	thereby causing malicious behaviour of HDT including denial of required treatments for the concerned patient. 
	\item \emph{Data injection}: In HDT, malicious data may be injected to alter the normal execution process. Such an attack can drive HDT into an unsafe state, making the system susceptible to data loss, DoS and data integrity attacks.
	\item \emph{Eavesdropping}: In HDT, eavesdropping attacks can be passive or active. An attacker quietly monitors message transmission and gathers useful information for the desired purpose in passive attacks. In contrast, active attacks occur when fraudulent nodes participate in communications, posing as legitimate ones to obtain important information, such as healthcare-related data and personal information, for misuse \cite{479}.
	
\end{enumerate}
\begin{enumerate}[a)]

\item \emph{Intrusion detection system}:
Intrusion detection system (IDS) is a critical security system aimed to manage attacks on networks while recognizing malicious actions in computer network traffics. 
IDS plays an imperative role in supporting data security by discovering, deciding and detecting unauthorized usage, duplication, modification and demolition of data and data frameworks \cite{493}. 
For instance, Chen et al. proposed effective algorithms for detecting low-rate DoS attacks in ZigBee networks of HDT in \cite{502}. They accomplish this by combining Hilbert-Huang transforms with trust evaluation approaches, thereby enhancing the security of ZigBee networks in HDT. Since most of the available IDS solutions can only detect attacks, but cannot measure the attack severity. To this end, Ramos et al. in \cite{503} proposed a node security quantification probabilistic model, based on the message security value and the damage level, to quantitatively evaluate the impact of compromised nodes on the integrity of data sent via 6LoWPAN networks. Additionally, traditional IDS-based security solutions usually suffer from a high false positive rate (FPR) and often need manual modifications, making it very difficult to scale when adopted in HDT \cite{368}. The recently proposed ML-based IDSs well address this issue. 
Schneble et al. in \cite{368} designed and implemented a massively distributed, FL-based intrusion detection system (FLIDS), which can achieve high accuracy, low FPR and communication cost, along with sufficient flexibility and scalability, making it suitable for the detection of attacks in HDT-enabled healthcare systems.
Thamilarasu et al. in \cite{370} developed a novel mobile agent-driven IDS for HDT using ML to secure the network of connected personal biomedical devices. The proposed IDS was hierarchical, autonomous and adopted regression algorithms to detect network level intrusion and anomalies in sensor data.
%
\item \emph{Cyber resilience}:
Cyber resilience is the ability to prevent, withstand and recover from cybersecurity incidents (i.e., cyber attacks), which is a critical enabler of trustworthy in the operation of HDT. Vulnerability tolerance is a fundamental requirement of cyber resilience in HDT. 
Exploitable HDT vulnerabilities are critical security threats to healthcare organizations and can affect massive users \cite{134}. 
One fundamental technology for cyber resilience in HDT is the software vulnerability detection technique. 
Existing conventional vulnerability detection techniques can be categorized into static and dynamic ones. Static techniques such as rule-based detection, code detection and symbolic execution, often result in many false positives, while dynamic techniques such as fuzz testing and taint analysis frequently suffer from low code coverage problems \cite{349}. To deal with these limitations, ML techniques have been widely adopted in software vulnerability detection. 
Lee et al. in \cite{352} proposed an improved ML-based static binary analysis technique to learn representations from code context. The obtained results showed that software vulnerabilities can be detected with an accuracy of 91$\%$ of the assembly code. Zhou et al. in \cite{353} introduced a graph neural network-based model for graph-level classification through learning on a rich set of code semantic representations to localize the vulnerable functions among the source codes.
Besides these, Zhang et al. in \cite{134} proposed a novel end-to-end scheme by adopting bidirectional long-short-term memory (LSTM) network with a self-attention mechanism for cyber resilience. The mechanism can explore a bi-directional relationship among some key codes, thereby recognizing potentially vulnerable functions in software projects for HDT. Similarly, a novel end-to-end vulnerability tolerance scheme was proposed in \cite{343} to recognize and fix HDT vulnerabilities. A new CodeBERT-based neural network was applied to better understand risky code while capturing cybersecurity semantics \cite{343}. The technique has been demonstrated to have great potential in the cyber resilience of HDT through extensive evaluations.
\end{enumerate}

\subsubsection{Privacy-Preserving Mechanism}
In HDT, network security is not sufficient to enhance operations, since medical data are also privacy-sensitive. Several common privacy threats in HDT are listed in the following. 

\begin{enumerate}[-]
	\item  Attackers may eavesdrop data on the public communication channel compromising data transmitted in plaintext.
	\item Since repeated keys are commonly used by users, it is possible that an attacker may break into the data storage server (e.g., cloud server) by brute force attacks based on key dictionaries collected from other data leakages.
	\item \emph{Honest-but-curious} provider of data storage or computing services (e.g., cloud server) may also try to compromise the data privacy of HDT.
\end{enumerate}

\begin{figure}[!t]
	\centering
	\includegraphics[width=0.95\columnwidth]{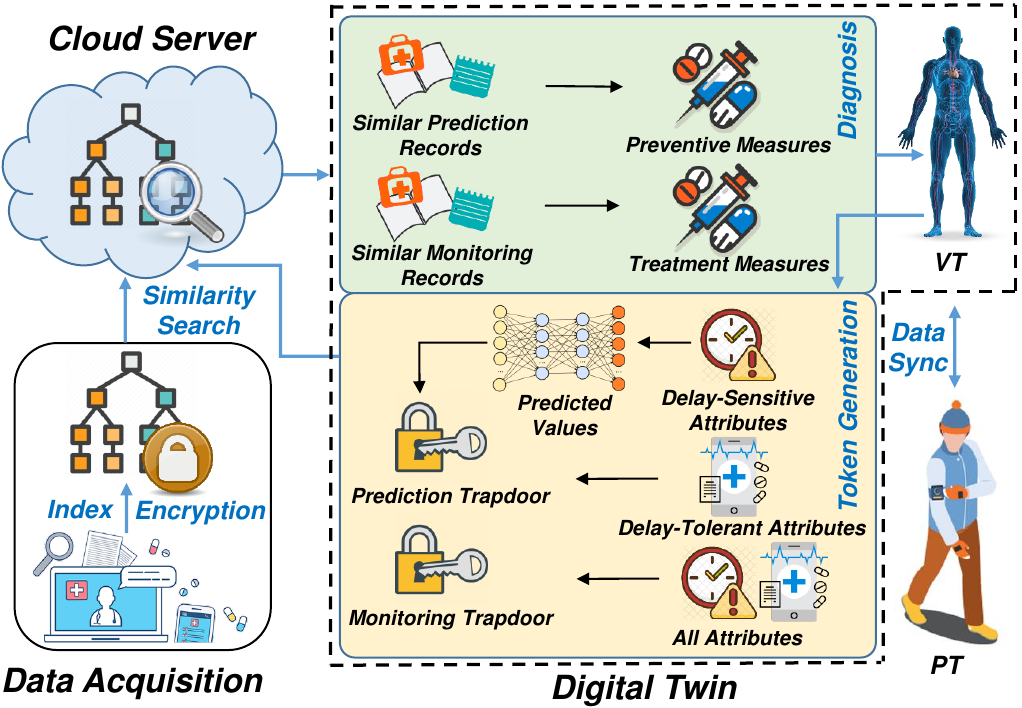} \\
	\caption{Cryptography-based privacy-preserving mechanism for HDT.}\label{query}
\end{figure}

As visually illustrated in Fig. \ref{query}, the main privacy-preserving mechanisms for HDT include cryptographic techniques, anonymization techniques, differential privacy and FL, which are reviewed in detail hereafter. 

\begin{enumerate}[a)]
\item \emph{Cryptography}:
Cryptography is a critical enabling technology for HDT to protect the privacy of personal health data in all segments. Cryptography techniques can be classified into several categories based on different criteria, such as symmetric-key cryptography, asymmetric-key cryptography, homomorphic encryption and quantum cryptography.

\begin{itemize}

\item \emph{Symmetric-key cryptography and asymmetric-key cryptography}:
Symmetric-key cryptography uses the same key for both encryption and decryption, while asymmetric-key cryptography uses a pair of keys, i.e., a public key for encryption and a private key for decryption. They can be used to secure the transmission and storage of sensitive healthcare-related data in HDT.
As a data source of HDT, the IoMT device introduced in Section \ref{SE3} monitors the health parameters of the human body and generates health-related data. These data must be kept safe from possible misuse and modification. The resource-constrained characteristic of IoMT devices, however, means it is very difficult to carry out data encryption with high-level cryptographic techniques. To overcome this bottleneck, efficient lightweight cryptographic techniques have been recently explored to support  IoMT devices \cite{373, 374}. 
Chaudhary et al. in \cite{373} proposed a novel block cipher-based technique by performing matrix rotation, XoR operation and expansion function to encrypt data. Noura et al. in \cite{374} designed a flexible lightweight cipher using a one-round simple cipher scheme with a dynamic key approach and introduced the concept of dynamic chaining and block mixing. 
Apart from the health data, the plaintext of query requests in HDT should also be kept secret from the \emph{honest-but-curious} data storage server (e.g., cloud server) and other devices. Zheng et al. in \cite{379} designed an efficient and privacy-preserving similarity query-based healthcare monitoring scheme, called PSim-DTH. Under such proposed scheme, the healthcare centers encrypt the data of each patient (e.g., EHR data) before outsourcing such data to the cloud. Then, the counterpart VT of a patient can launch similarity range queries to the cloud server. The cloud server is expected to return data records that are similar to the patient, as the query results, to the VT. Additionally, to speed up similarity query efficiency while guaranteeing data privacy, the authors adopted a partition-based tree (PB-tree) to index the healthcare data and introduced matrix encryption to present a privacy-preserving PB-tree-based similarity range query (PSRQ) algorithm.

\item Homomorphic encryption: This is a special type of encryption that allows computations to be performed on encrypted data without first decrypting it. The homomorphic encryption has been utilized in healthcare system for protecting the security and privacy of healthcare-related data, and such paradigm can be also applied in segments of HDT, such as in the communication layer, computation and data management layer. For example, Zhang et al. \cite{470} proposed a FL mechanism for deep learning of medical models in IoT-based healthcare systems. They further employed homomorphic encryption to protect local models from inference attacks on private healthcare-related data. Shaikh et al. \cite{471} utilized the fully homomorphic encryption techniques to secure the ECG signals, which helps the medical provider to record ECG signals confidentially and to prevent mistreatment.

\item Post-quantum cryptography: Post-quantum cryptography can enable the HDT by providing a secure method for protecting the privacy-sensitive healthcare-related data that is exchanged between PTs and VTs. With the emergence of quantum computing, current cryptographic methods that are used to secure sensitive data, such as EHRs, could be vulnerable to attacks in the future. Fortunately, the post-quantum cryptography is developed to resist quantum computers and quantum computing-based attacks. This means that post-quantum cryptography can provide a high level of security and privacy for HDT in PH applications. For example, Mirtskhulava et al. \cite{472} proposed a blockchain scheme with hash-based post-quantum digital signatures for securing EHRs. Xu et al. \cite{473} proposed a post-quantum public-key searchable encryption scheme on blockchain (PPSEB) for E-healthcare scenario to enable data sharing while ensuring security and privacy. PPSEB employed a lattice-based cryptographic primitive to ensure the security of the search process and achieve the forward security to avoid key leakage of medical information.
\end{itemize}
\item \emph{Anonymization}:
Anonymization captures the process of removing identity information from health data to mitigate the risks of identifying the actual source or owner of such health records. Majeed in \cite{383} proposed an anonymization scheme to enhance the data privacy of EHR. The proposed scheme aimed at preventing identity disclosure even in the presence of pertinent background knowledge by adopting a fixed interval approach, which classified the quasi-identifiers of the EHR data in fixed intervals, while replacing the original values with their averages. 
Aminifar et al. in \cite{384} investigated the process of anonymizing data in a health data sharing application to address the problem associated with record-linkage and attribute-linkage attack models. To ensure anonymization, the authors studied anonymization as a constrained optimization problem, where k-anonymity, l-diversity and t-closeness privacy models were jointly studied. 
Recently, ML has been demonstrated as a useful tool to enhance the anonymization of health-related data. A common example of an ML-based anonymization solution in the literature is generative adversarial networks (GANs)-based anonymization. Angulo et al. in \cite{380} proposed an HDT architecture to support patients with lung cancer. The proposed architecture adopted GANs to ensure anonymization of healthcare data (e.g., CT images of a patient). Through GANs, fake health data (e.g., thyroid-related and cardiogram data) were also generated in \cite{381} by converting the original raw data, including static data and dynamic data, into images to facilitate their usage for GANs anonymization process. 
\item \emph{FL}:
FL is a distributive AI paradigm which has opened up new opportunities to ensure privacy in HDT by ensuring that raw data of participants are not transmitted during any data sharing process. 
The centralized AI paradigm has the potential to suffer from privacy leakage because sensitive health data are transmitted through public networks and processed by a \emph{honest-but-curious} cloud server. In contrast to centralized AI, FL promotes distributed learning on end devices, where only gradients are transmitted to the central global model, thus preserving the privacy \cite{386}. Note that we have already surveyed the applications of FL in HDT in section \ref{EC-CLO}, and therefore the details are not repeated again for conciseness. 
\end{enumerate}

\subsubsection{Distributed Ledger Technology}
Distributed ledger technology (DLT) is a type of digital database that allows for the secure and transparent recording and storage of data across multiple nodes or participants in a network. Unlike traditional databases, which are typically centralized and controlled by a single entity, DLT is decentralized and distributed, with data stored on multiple nodes in a network. One of the most well-known technologies of DLT is blockchain, and hereafter, we introduce the applications of blockchain in HDT in detail.

Blockchain is widely recognized as a distributed and decentralized peer-to-peer (P2P) data storage mechanism, where transactions are validated and appended to the chain if a consensus is reached among a group of validators. Once the information is stored in the blockchain, it cannot be modified \cite{491, 492, 494}. As a result, blockchain becomes a suitable technology to enhance security in HDT-enabled PH applications and systems. 

\begin{figure*}[!t]
	\centering
	\includegraphics[width=0.93\textwidth]{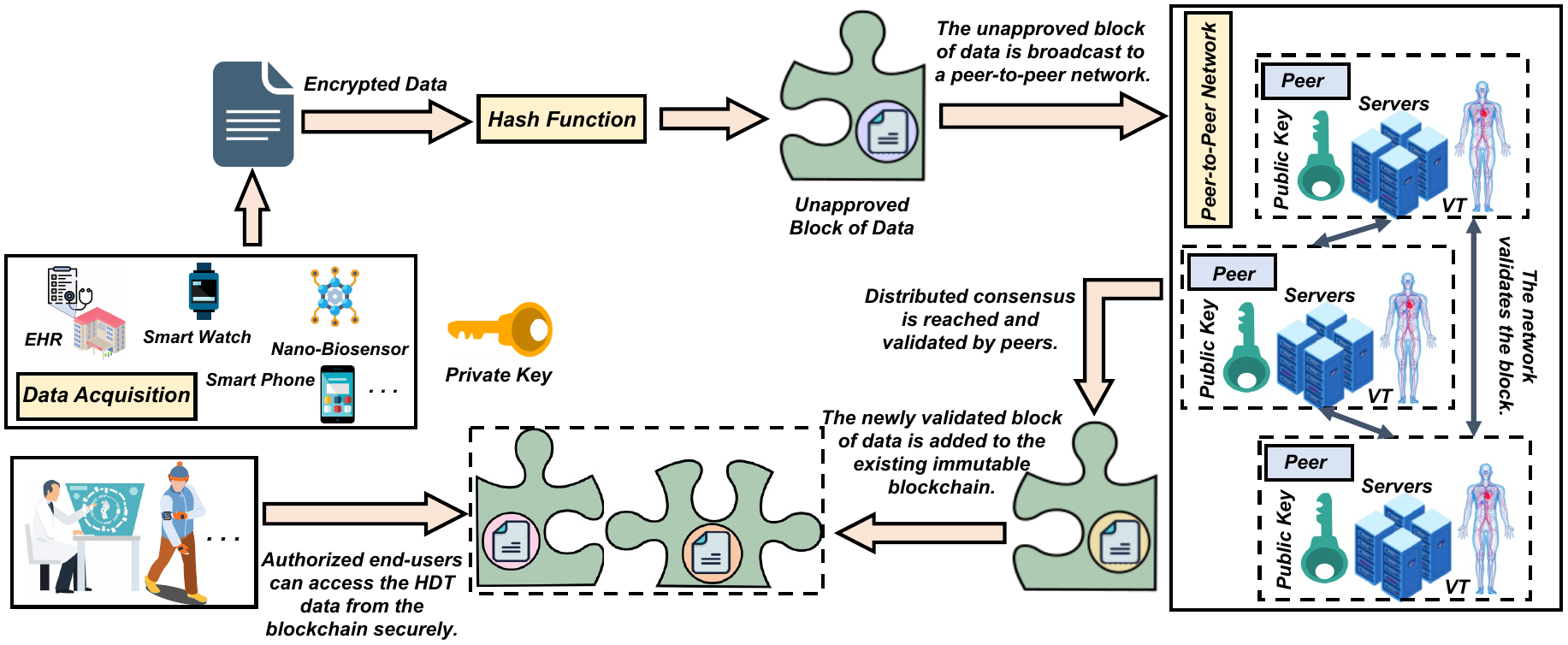} \\
	\caption{Blockchain for data sharing in HDT.}\label{BL_1}
\end{figure*}

\begin{figure}[!t]
	\centering
	\includegraphics[width=0.89\columnwidth]{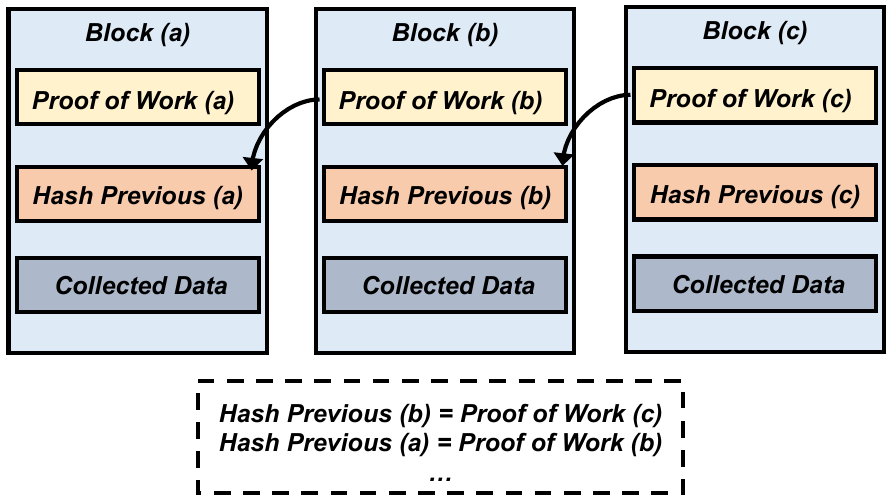} \\
	\caption{Structure of the Blockchain. } \label{BL_2}
\end{figure}

The procedure of a blockchain-enabled data sharing scheme for HDT is presented in Fig. \ref{BL_1}. In general, each end-user possesses two keys: 1) a public key for encryption and 2) a private key for decryption, as in asymmetric cryptography. From the blockchain perspective, the private key is used to sign the blockchain data, and the public key represents the unique identity. At the initial stage, a typical end-user (e.g, a smartphone or a medical institution) signs a set of collected data cryptographically using its private key and generates a unique hash to form an unapproved data block. This hash is unique to such a data block and changes if the data in the block is changed. After the generation of an unapproved data block, such a block is broadcasted across the network to its peers (a peer-to-peer network). Once the peers receive the signed data block, each peer validates the block and disseminates it over the network. All the involved end-users mutually validate the collected data to reach a consensus. When consensus is reached, such a block is hash-matched with the previous block in the blockchain and then appended to the blockchain \cite{495}. As shown in Fig. \ref{BL_2}, the blockchain structure generally consists of a group of linked blocks. Each block in a blockchain architecture contains a set of data and can be identified using a hash function, consisting of the hash of the previous block, a proof of work (derived by a special peer (called miner) in the validation process) and the collected data \cite{496}. 

\begin{enumerate}[a)]
	
\item \emph{Storing health data}:
When adopted in HDT systems, blockchain can ensure the storage of health data in an immutable, secure and reliable manner. Zhao et al. in \cite{338} proposed a blockchain-enabled health application to enhance the protection and storage of physiological signals collected through BSNs. All blockchain nodes participate in ensuring a reliable system through consensus mechanism and digital signature while using hash chain technology to maintain the proposed health blockchain. Similarly, Yue et al. in \cite{339} proposed a healthcare data gateway (HDG)-centric healthcare architecture for enabling each patient to manage and control their medical data securely. The proposed architecture is composed of three layers, i.e., data storage, data management and data usage layers. Despite the ability of blockchain to ensure security and privacy when adopted for storage, HDT applications are known to generate a massive volume of data, which cannot be stored on-chain. As a result, efforts are now focussing on how to accommodate massive storage with blockchain without compromising the overall performance.

\item \emph{Data access control}:
Rather than adopting blockchain as a database, blockchain could be better used to ensure data access control, thereby providing a secure data sharing service. Zhang et al. in \cite{336} proposed a double blockchain-enabled secure storage and sharing framework for EHR data. To ensure privacy, the EHR data were first encrypted and stored in the cloud. After this, the storage blockchain was used to store a complete EHR data index, while the shared blockchain stored the index of the shared part of the EHR data. Users with different attributes can make requests to different blockchains to share different parts according to their permissions. A novel decentralized health architecture was proposed in \cite{115} for a cooperative hospital network by integrating MEC and blockchain technologies to improve data offloading and sharing while ensuring users' QoS and security awareness. In this work, health data were encrypted and stored in an edge-enabled interplanetary file system storage platform. 
Furthermore, blockchain and smart contracts were adopted to ensure auditability. 
For instance, Liu et al in \cite{416} proposed a blockchain-based privacy-preserving data sharing scheme for EHR data, where the original EHR data were encrypted and stored securely in the cloud while the indexes were reserved in a tamper-proof consortium blockchain. Besides, the authors implemented smart contracts in the consortium blockchain to secure the data accessed, while each request and access activity were similarly recorded for future auditing. 
\end{enumerate}

\subsection{Lessons Learned}\label{SE6_4}
Sheer volume of data are constantly being generated and exchanged in HDT between the PTs and their corresponding VTs. This presents significant challenges in data management. Data generated in HDT exhibit characteristics of heterogeneity, scalability, and with high noise, and therefore require pre-processing, which includes data cleaning, data reduction, and data fusion. After the process of data pre-processing, those data require powerful data storage framework to store in robust and efficient ways. Additionally, data security and privacy are also crucial aspects of data management for HDT, which can be enabled through key technologies such as cybersecurity, privacy-preserving mechanisms and blockchain.

However, there are several opening issues that have not been perfectly resolved to date. For example, the standardization and interoperability of HDT data need to be developed, as the lack of these may significantly degrade the data value. This motivates researchers to further study the harmonization of data formats, ontologies, and protocols to enable effective data sharing and integration for HDT in PH applications.

%

\section{Key Technologies for Data Analysis and Decision Making Layer} \label{SE7}

\begin{figure}[!t]
	\centering
	\includegraphics[width=0.99\columnwidth]{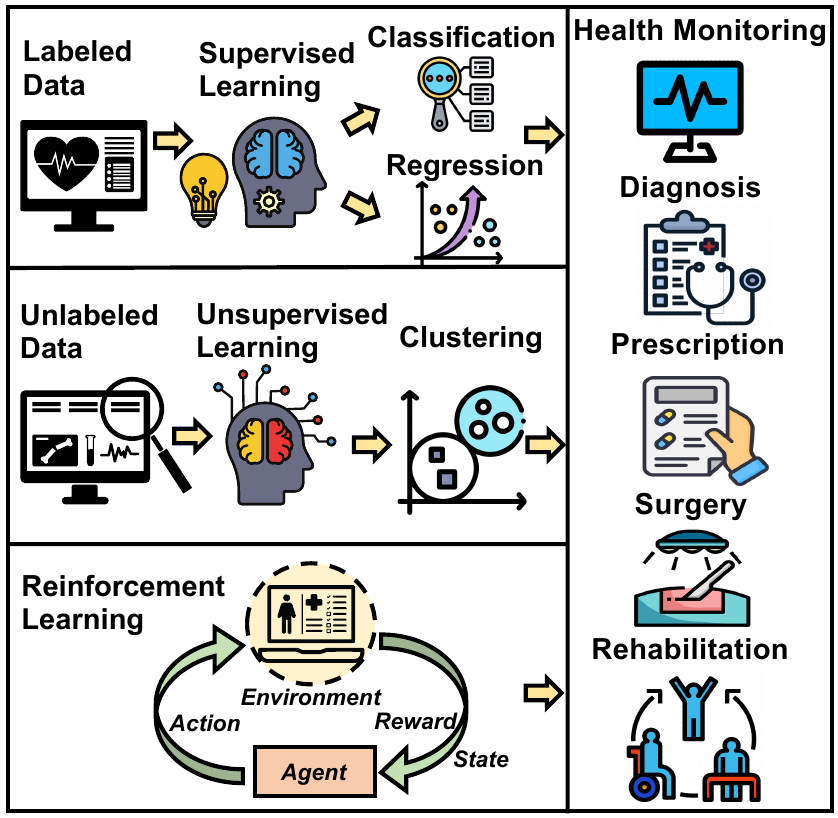} \\
	\caption{AI for data analysis and decision making in HDT applications. } \label{AI}
\end{figure}

AI is one of the main fueling technologies to support the design and implementation of HDT. Aside from its usefulness for model evolution, AI remains a significant tool for big data processing. AI is capable of empowering machines with human-like intelligence to support the mining of underlying valuable information in big data. As shown in Fig. \ref{AI}, in this section, we explore the application of three main AI categories, i.e., supervised learning with labeled data (e.g., KNN, SVM), unsupervised learning with unlabeled data (e.g., K-Means, GAN) and reinforcement learning (e.g., Q-learning) in HDT-enabled PH applications and systems. We demonstrate how AI can energize key functions of HDT, such as health monitoring and diagnosis, prescription, surgery and rehabilitation.

In this section, we discuss AI-powered data analysis and decision-making for HDT in PH applications, including diagnosis, prescription, surgery, and rehabilitation. Specifically, we survey the role of AI in HDT applied to personalized monitoring and diagnosis in Section \ref{SE7_1}. Then, in Section \ref{SE7_2}, we focus on surveying AI-enabled HDT solutions for personalized prescriptions. Furthermore, we discuss the applications of HDT in surgery and rehabilitation with the assistance of AI in Sections \ref{SE7_3} and \ref{SE7_4}, respectively. Finally, we summarize this section and propose several opening issues that should be considered in the implementation of data analysis and decision-making for HDT in Section \ref{SE7_5}.

\subsection{Diagnosis} \label{SE7_1}
One of the key requirements in PH service is personalized monitoring and diagnosis.
By leveraging AI, HDT can provide personalized and precise diagnosis services \cite{509, 515}. For instance, an ML-based human heart VT was proposed in \cite{133} for the detection of heart diseases. Implantable devices, such as pacemakers and implantable cardioverter defibrillators, were used to capture data related to heart conditions. The captured data were then transmitted to the cloud to support the creation and storage of heart VT. A decision tree based on an ML algorithm, was used to categorize the conditions of patients according to real-time data obtained from the VT as well as the previously stored health data in the cloud. The decision tree classifier achieved an accuracy of 79$\%$, which was comparable with the traditional medical treatment.
Martinez et al. in \cite{95} designed an edge-assisted cardio twin platform for real-time detection of IHD. The system adopted CNN  to classify myocardial infractions (a type of IHD) and non-myocardial infractions. The adopted CNN framework was able to generate features from the ECG and performs the classification task. The system achieved an accuracy of 85.77$\%$ while achieving timeliness of 4.84 ms to complete the classification of each sample.
Similarly, an HDT solution was proposed in \cite{132} for patients in an IoT context-aware environment. The HDT solution was necessary to aid the monitoring of real-time health status as well as detection of body metrics anomalies by building a novel ECG heart rhythms classifier model that diagnoses heart disease and detects heart problems. The results showed that LSTM achieves the best performance in terms of accuracy, precision, recall and F1 score.
Another typical example is the diagnosis of the aortic abdominal aneurysm (AAA), a increasingly growing dilation of the aorta with a risk of potentially lethal rupture. Because of the high mortality rate of around 80$\%$, any successful treatment of AAA usually depends on how early it was detected. In \cite{135}, the authors designed a cardiovascular VT for the detection and diagnosis of AAA, in which HDT was modelled through the adoption of the LSTM-based inverse analysis. The CNN classifier was trained to extract features from the blood pressure waveform to detect AAA. 
The results showed that the accuracy of the customized CNN in detecting AAA achieves 99.91$\%$, while the classification of its severity can achieve 97.79$\%$, which are surprisingly higher than those by traditional clinics. 


Compared to conventional DT, the provision of explanations or justifications to verify the reliability of machine-enabled predictions and decisions is particularly crucial in the field of diagnosis. This is because the predictions or decisions made by machines can directly impact the safety and well-being of individuals, and reliable explanations or justifications are essential for ensuring the credibility and confidence of the results \cite{512}. To this end, a new concept, called explainable AI, has been proposed recently \cite{137}. Explainable AI is capable of providing decision outputs based on the set of inputs while also providing a set of supporting evidence. This technique has been integrated into HDT. For example, in the HDT-based personalized elderly type 2 diabetes management framework developed by Thamotharan et al. in \cite{512}, the local interpretable model-agnostic explanations (LIME) was used to explain the factors that may lead to hyper and hypo events in a patient, thereby helping personalization and improving patient behavior.
 Besides, Pitroda et al. in \cite{148} developed a lung disease identification model using the explainable AI technique. A customized CNN with the ability to obtain reasonable performance in terms of classification accuracy was trained. 
 Then, a layer-wise relevance propagation (LRP)-based method was proposed to quantitatively interpret the performance of the CNN model.

\subsection{Prescription}\label{SE7_2}
Here, we focus on surveying related AI-enabled HDT solutions for personalized prescriptions, or also called personalized treatment planning.

With HDT, prescriptions can be first tested on any counterpart VT, located in the virtual environment, to understand their performance, and then the resulted prescription with the best performance can in turn be given to the corresponding patient in the physical environment. 
For example, a high-resolution HDT of an individual patient was constructed in \cite{146} to improve the personalized prescription process. To achieve this, unlimited copies of the VT of a patient (which includes all molecular, phenotypic and environmental factors relevant to disease mechanisms in such a patient)  were first created. Different medications were then given to each copy of the VT to computationally evaluate the medication with the best performance. The patient was subsequently treated with the optimal medication. Similarly, the authors in \cite{171} developed an HDT solution to detect sepsis. The solution was explicitly designed following a causal AI approach to predict possible responses to any specific treatment during the first 24 hours. Specifically, directed acyclic graPH (DAG), also known as the Bayesian networks, were used to define the casual relationship among organ systems and actual treatments.
In \cite{174}, another HDT framework was designed to determine optimal treatment decisions using survival and toxicity metrics. The system was responsible for predicting treatment outcomes in head and neck cancer patients relying on deep-Q-learning (DQN) \cite{487}. The results showed that the mean and median accuracies can reach 87.09$\%$ and 90.85$\%$, respectively, while the survival rate can increase by 3.73$\%$. Given the prediction accuracy and predicted improvement in medically relevant outcomes obtained by this approach, the authors concluded that HDT has the potential to greatly help physicians to determine the optimal course of treatment and assess the outcomes of it.

Predicting the course of any disease will help medical personnel when deciding the most appropriate treatments for each patient. 
HDT has been proven to be a useful approach for simulating disease progression due to its adoption of various AI algorithms. Allen et al. in \cite{138} adopted variational autoencoder ML methods when creating HDT to forecast the trajectories of multiple clinical measures in patients experiencing an ischemic stroke, a disease reported as the second most common cause of mortality globally \cite{180}. 
The proposed method was demonstrated to be capable of accurately capturing the cross-sectional and longitudinal properties of real patient data. Moreover, HDT has proven to be a possible solution for chronic neurological conditions \cite{145}. Unlike most ML-based solutions for disease progression prediction, where only a single endpoint can be predicted, Fisher et al. in \cite{139} trained an unsupervised ML model, called conditionally restricted Boltzmann machine (CRBM), for individualized forecasting of entire patient trajectories with Alzheimer's disease. Synthetic patient data generated by the CRBM precisely reflected the means, standard deviations and correlations of each variable over time, while the results also showed that synthetic data cannot be distinguished from actual data through logistic regression. 
Another potential application is for the cancer treatment. Although Stereotactic radiotherapy (SRT) is an effective treatment modality for cancer patients with spinal metastasis (SM), it may weaken the bone tissue and further increase the risk of vertebral fracture (VF). Therefore, predicting the risk of VF in SM cancer patients is crucial for making better treatment strategies. The authors in \cite{177} designed an AI-assisted framework, called ReconGAN, for creating a VT of the human vertebra and predicting the risk of VF. The VF response of the final vertebra model was simulated using a continuum damage model under compression and flexion loading conditions. The authors in \cite{178} designed an HDT framework based on a modular AI-aid system, which was used to model the whole human body and provide a panoramic view of current and future pathophysiological conditions.

\subsection{Surgery} \label{SE7_3}

According to WHO, more than 2.5$\%$ of patients die during surgery or after surgery. This high mortality rate can be attributed to in-surgery and post-surgery reactions, low precision of the used technologies as well as lack of suitable experience of surgeons \cite{185}. Fortunately, AI-enabled HDT can be applied in preoperative, intraoperative, and postoperative management for improve surgical accuracy and precision \cite{507}.

AI-assisted preoperative planning can be an essential method to guarantee surgical success. 
By the traditional way, surgeons prepare themselves for a surgical procedure by looking at medical records and images, such as CT and ultrasound of the concerned patient. This medical imaging-based preoperative planning routine includes anatomical classification, detection, segmentation and registration \cite{186}. To ensure more accurate identification and classification of colon lesions and anatomical landmarks in colonoscopy images, Jheng et al. in \cite{188} developed a CNN-based algorithm to classify benign or malignant lesions of thyroid cancer in ultrasound images.
Rubinstein et al. in \cite{194} applied an unsupervised learning approach to detect and localize prostate cancer foci in dynamic positron emission tomography (PET) images. To achieve this, they trained a deeply stacked convolutional autoencoder to extract statistical, kinetic biological and deep features from 4D PET images. Torosdagli et al. in \cite{197} developed a fully convolutional DenseNet-based automated image analysis software for mandible segmentation of cone-beam computed tomography (CBCT) images.
A UDS-Net-based method was proposed in \cite{198} to segment teeth from dental CBCT images. The proposed method consists of a U-Net with a dense block, comprising multiple densely connected convolutional layers and spatial dropout. Furthermore, a novel 3D deep supervised densely network (3D-DSD Net) was proposed in \cite{199}. The proposed 3D-DSD Net consists of a U-Net with a dense block and additional skip connections, to execute CT data segmentation tasks of the small organs in the temporal bone.

Besides all above, several HDT-based preoperative planning platforms have also been developed. For instance, an HDT-based remote surgical rehearsal platform was introduced in \cite{316} to improve surgical simulation experiences. This work proposed a novel robust auxiliary classifier GAN (rAC-GAN)-based intelligent prediction model to predict the pathology of lung cancer with pulmonary embolism, and combined mixed reality (MR) to project surgical navigation images. Furthermore, an HDT-based laparoscopic surgery assisting tool integrating AR and ML was developed in \cite{354}.

AI can enable surgical robots to achieve superhuman performance. 
Learning from demonstration (LfD) is a prevalent paradigm for allowing robots to autonomously and intelligently perform tasks using learned strategies \cite{204}. The paradigm is beneficial for surgical procedures, where surgical robots can intelligently execute specific tasks or motions by learning from surgeons' demonstrations without tedious programming procedures. For instance, LfD has been shown to be a promising solution for aiding surgical robotic systems to automatically execute soft tissue manipulation \cite{205}.
Human-robot interaction (HRI) is another essential part of surgical robotic systems. With the assistance of AI, the HRI enables surgeons to control the surgical robot with touchless manipulation \cite{516}. A dense CNN-based model was developed in \cite{207} for 9-direction/36-direction gaze estimation. Then, with the help of the proposed model, doctors can easily control the robot by specifying the starting point and ending point of the surgical robot using eye gazing.
 Fujii et al. in \cite{208}  performed real-time gaze gesture recognition with the hidden Markov model (HMM), allowing the laparoscope to pan, tilt and zoom during surgery, whilst immune to aberrant or unintentional eye movements.

\subsection{Rehabilitation}\label{SE7_4}
Due to the increasing aging population, rehabilitation of diseases like stroke poses a huge healthcare burden, resulting in a demand for new rehabilitation methods. One of the most promising solutions for this problem is HDT with AI assistance \cite{221}.

AI could assist HDT in predicting patient movements to improve the performances of lower and upper limb rehabilitation. For the upper limb rehabilitation, a multi-stream LSTM duelling-based motion prediction model was proposed in \cite{224}. Specifically, the multi-stream LSTM duelling was employed to predict the trajectories with a multi-joint motion of the human arm and then utilized the predicted angles as input to control the trajectory of the exoskeleton robot, thereby ensuring synchronization between the robotic and human arms. For lower limb rehabilitation, Alekseyev et al. \cite{524} have developed an HDT-based rehabilitation system, featuring two measuring modules located in the heel area, relying on a micromechanical sensor with a three-axis accelerometer and a gyroscope. The measurement data is analyzed using AI algorithms to assess the level of recovery in walking skills.

AI can assist HDT in providing therapists with quantitative analysis of patients' rehabilitation status to improve practices \cite{525}. As an example, Lee et al. in \cite{227} designed an intelligent decision support system for stroke rehabilitation assessments. The design was able to automatically identify salient features of assessment using RL to assess the quality of motion while generating a summarized patient-specific analysis as an explanation of its prediction. A novel rehabilitation assessment paradigm was proposed in \cite{228}, where a domain expert and an AI-based system were integrated to collaborate in executing complex decision-making tasks, such as stroke rehabilitation assessment. The results presented in this work showed that accuracy of decision-making can be significantly increased via AI assistance. 

\subsection{Lessons Learned}\label{SE7_5}
AI is widely known as the key technology for data analysis and decision making for HDT in PH applications. Generally, there are three categories of AI algorithms, i.e., supervised learning, unsupervised learning, and reinforcement learning.
These powerful algorithms assist HDT to analyze and mine valuable information from HDT data, enabling intelligent diagnosis, prescription, surgery, and rehabilitation.

However, there are still some opening issues that need to be considered. First, although explainable AI has gained significant momentum, its full-fledged explainability in the healthcare domain can hardly be achieved, and the reliability concern can never be neglected.
Second, the training of AI algorithms for HDT may incur significantly high computing costs, and therefore low-complexity AI models without compromising accuracy are highly desired for HDT.

\section{Future Research Directions} \label{SE8}
The adoption of HDT for PH is disruptive and revolutionary. It can drive precise and effective interventions tailored to every user by duplicating a high-fidelity and interactive digital representation of the real human body. Despite these advantages, many technical issues hinder the implementation of HDT, and thus can be seen as future research directions.

\subsection{Data Scarcity in HDT}
Data is the bedrock of HDT, but it has also become a major bottleneck in the development of HDT. To create hyper-realistic and hyper-intelligent testbeds, training datasets for HDT are collected not only from PT, but also from many medical institutions that possess larger amounts and more comprehensive historical data. However, the data and information in current medical systems are often fragmented and severely segmented, resulting in data heterogeneity and bias. Additionally, different medical institutions may use various types of equipment, each with its own standards, resulting in the absence of standardized data formats or interfaces. Medical data is also relatively sensitive, and to protect patient privacy, most medical institutions are reluctant to share their data. These issues indicate the urgent need to develop a unified and secure data management framework to break down data silos, or to leverage artificial intelligence-generated content (AIGC) to synthesize realistic datasets with a small amount of real-world data, in order to overcome the data scarcity challenges faced by HDT.

\emph{Use-case}: In situations where a patient is afflicted by a rare disease, the scarcity of adequate data required to create a high-fidelity virtual model of the disease would have a significant impact on the treatment process. For instance, doctors may be unable to prescribe personalized medication based on the versatile testbed function of HDT. In such scenario, rare disease-related datasets from numerous medical institutions become vital, and data management systems must be interoperable to enable their seamless access. Moreover, synthesizing the rare disease-related datasets via AIGC models may be an alternative solution.

\subsection{Mobile HDT}
HDT implementation must consider PTs' complex mobility patterns in the physical environment. These patterns can be categorized into human positional and postural mobility. Both positional mobility, like a PT moving from indoor to outdoor, and postural mobility, like lying and sitting, may cause the PT-VT connectivity and service interruptions. These factors can impact the reliability, stability, security, and cost of the implementation of HDT in PH applications. To guarantee seamless and pervasive connectivity between a PT and its corresponding VT during mobility, the location of each VT must depend on the current location of its paired PT. With this, a VT migration policy is essential to realize high-fidelity HDT frameworks. Possible solutions to support mobility in HDT include the adoption of multi-hop techniques as well as edge-cloud computing-enabled HDT migration techniques. The migration of HDT, however, raises other imperative challenges including handover issues, energy efficiency, security and privacy, latency, etc., that need to be considered to ensure its successful implementation. Thus, developing a novel framework supporting the migration of HDT as well as formulating key optimization parameters remains as an open problem.

\emph{Use-case}: When a PT (representing a patient) is at home and is equipped with various biomedical devices to collect its real-time health data and upload to a server for updating the VT, the server should be located near the home so as to be fast-responsive. However, the PT is not confined to be at home, and may move around to various places such as its workplace or social events, or even on fast moving vehicles. Under these circumstances, the VT should migrate to nearby servers along with the PT as it moves, for ensuring seamless and real-time connectivity between the PT and its corresponding VT.

\subsection{Federated HDT in the Cloud/Edge Network}
A corresponding VT of any typical PT may need to be replicated and distributed in multiple computing servers such as cloud-edge servers to achieve the federated HDT. This is important since existing networks are known to be resource-constrained in terms of storage, power, computing and high-speed memory, making it difficult to maintain VT on a single server. Similarly, massive and intensive computing tasks are required for the construction and evolution of VT. This is necessary to alleviate unnecessary performance bottlenecks for HDT. Replicating VT across distributed networks can enhance a collaborative data analysis process, thereby eliminating resource constraint issues associated with a single-server computation scheme. Furthermore, VT replication on multiple servers can ensure fault tolerance. In a single-server mechanism, a server failure will hinder the required seamless PT-VT connectivity, making it difficult to achieve synchronization at any time. PTs in federated HDT can maintain synchronization even in the presence of one or more server failures by switching to another functioning server containing its VT.
	To achieve federated HDT, many research questions must be answered. For example, a) how can we replicate and distribute VTs across multiple servers in the network; b) what are the methods to achieve synchronization among these distributed VTs; c) what are the possible collaboration mechanisms for VTs; and d) how can we ensure reliable connectivity between any PT and its counterpart VTs; e) how to adaptively orchestrate the computing and networking resource for heterogeneous tasks in the federated HDT, which may be located in different places and have different QoS requirements. These and many other issues related to federated HDT require further investigations.

\emph{Use-case}: Suppose an HDT is collecting real-time health data from multiple data sources for implementing HDT construction, evolution and data analysis, etc. The amount of data may be too large to be stored and processed on a single server with constrained resources, while the HDT is expected to keep seamless PT-VT connectivity in case of a single server failure. In this scenario, the federated HDT can be adopted by replicating and distributing the VT of the PT across multiple cloud and edge servers. Each server can store a portion of the VT and the data will be synchronized among the servers in real-time. By this way, the PT can always keep up-to-date and available, even if one of the servers collapses. Besides, the replication of the VT across multiple servers will allow a collaborative data analysis process, accelerating the speed of the data analysis process.

\subsection{Interoperable Management of Subsystems in HDT}
Human body is an extremely complex system composed of many tightly-coupled subsystems, including the circulatory, respiratory, digestive and nervous systems. Each of these systems has its own unique functions, but they also closely collaborate and influence each other. Similarly, the orchestration of an individual's health data management layer should follow these subsystems, with each supporting its own corresponding DT, such as the circulatory system DT. The data management of these DTs should be designed to support the mutually sharing of information, while also ensuring interoperability on semantic, data, and other levels. This will enable fully integrated and optimized personalized healthcare services. Promising solutions may be integrating semantic interoperability based on ontologies and model transformation-based approaches.

\emph{Use-case}: Consider applying HDT to personalized monitoring, and focus on the collaboration of the respiratory and circulatory system DTs. When a PT is exercising, the respiratory system needs to increase efficiency to deliver more oxygen to lungs, while the circulatory system needs to pump more blood to muscles, resulting in a feedback loop between these two systems. In its HDT, if the respiratory and circulatory system DTs cannot achieve data level interoperability, it would fail to accurately modeling the body's physiological processes. Therefore, the management layer needs to ensure that there is seamless coordination and interoperability among each of the subsystems' corresponding DTs.

\subsection{Effective and Efficient Interface Design}
Interface design for the networking architecture supporting HDT in PH is critical to the integration of all layers across the data acquisition layer to the data analysis and decision making layer. It is expected to be human-centered, multi-modal, secure, efficient, scalable, interactive and standardized.
Firstly, the interface design for data acquisition should prioritize the comfort of patients and ensure seamless integration among multiple heterogeneous medical sensing hardware and communication approaches. For example, a skin-integrated wireless haptic interface \cite{535} can be designed as a comfortable and user-friendly way to enable data interaction between a PT-VT pair.
Secondly, the interface design should consider multi-modal input and output design, including audio, gesture, video signals, and the ability to handle large volumes of data while maintaining high levels of accuracy and reliability.
Thirdly, the interface design should consider efficiency and scalability, especially in handling massive and complex computing tasks in PH services, which usually involves collaborations among multiple parts of ``human''. This requires the interface design to consider parallel processing and distributed computing for optimizing the performance of the computing layer in HDT.
Fifthly, the interface between the data analysis and decision-making layer and users be intuitive, such that users are able to interpret and perceive feedback of VTs easily (e.g., drug response and future disease progression).
Last but not least, designing a standardized interface for HDT is important to simplify the management of resources required for HDT implementation. The standardized interface can provide a unified way for users to access various resources, improving the efficiency and effectiveness of PH service delivery.

\emph{Use-case}: Suppose that there is a scenario in which a doctor uses a patient's HDT as a testbed for prescribing personalized medications. The doctor inputs a virtual candidate drug into the VT, and the VT visualizes the drug-disease interactions through the carefully designed user interface. By doing so, the doctor can vividly and intuitively understand the performance of the tested drug through the interactive interface, making informed decisions with ease.

\subsection{Intelligent Blockchain for HDT}
Blockchain is an important enabling technology of HDT to protect the security and privacy of users. However, blockchain-enabled systems can suffer from high latency due to the common adoption of complicated consensus mechanisms necessary to validate each transaction. Thus, intelligent blockchain solutions are required to ensure security and privacy, while satisfying the ultra-low latency requirement in HDT. Moreover, such an intelligent blockchain should be implemented in both the virtual and physical environments to ensure proper validation of every activity between each PT-VT pair as well as activities among VTs in the virtual environment. 

\emph{Use-case}: Imagine a scenario where medical staff are intervening in a patient in an emergency through HDT, where medical staff need to access massive data in blockchain-enabled HDT with frequent and timely interactions. However, the traditional blockchain may hinder this process due to its high latency in complicated consensus mechanisms. For this problem, the intelligent blockchain with lightweight consensus mechanism can be adopted to provide fast-responsive validation process without comprising security and privacy.

\subsection{Full-Fledged Explainable AI for HDT}
The role of AI in HDT should not be underestimated. Particularly, decision explanation is a crucial factor in HDT. While AI has shown its strong muscle for providing reliable data analysis in the existing literature, it is generally referred to as a black box model, due to its inability to offer details and understandable explanations for its decisions. This explainability feature is, however, essential in the healthcare domain to ensure trust and confidence in decisions. In other words, transparency in AI-based data analysis is required. To support medical personnel and improve their efficiency, HDT should provide full-fledged explainable recommendations. This will assist medical professionals in reaching safe and reliable decisions to provide PH services to each individual. Although explainable AI has recently started to attract some attentions \cite{137}, more efforts are needed to investigate how this concept can be integrated to facilitate decision makings in HDT systems and applications. The current explainable AI solutions are also lacking in precision and detail, making the need for more in-depth research in this area a necessity. 

\emph{Use-case}: Take the remote monitoring and management of a patient with chronic conditions by HDT as an example, where HDT with AI can alert the patient or its healthcare professionals when the health status is abnormal for intervention. The lack of transparency in AI-based decision-making can make it difficult to understand the reasons behind the recommendations, and thereby results in reliability concerns. Therefore, the integration of full-fledged explainable AI with HDT can help improve the transparency and trustworthiness of AI-based recommendations, so that healthcare professionals can follow the reasons to infer the pathology. Ultimately, the integration of the full-fledged explainable AI will help medical professionals make more informed decisions and provide better care to patients with chronic conditions, even in remote settings.

\subsection{Generalized AI for HDT}
HDT-enabled PH systems may integrate a variety of AI algorithms such as KNN and SVM that need to be trained by labelled data. However, most health-related data in HDT are unlabelled. To improve performance, new AI algorithms that can efficiently perform training using unlabeled data are required to facilitate real-time and reliable decisions in HDT. Furthermore, the development of training-specific AI models for each task is time-consuming. The development of general artificial intelligence (GAI) through transfer learning \cite{413} and meta-learning \cite{414} opens several opportunities of alleviating the need for redundant computation.

\emph{Use-case}: Consider that a patient with a chronic disease, such as diabetes, is being monitored through its HDT. The HDT collects a large amount of health-related data from various sources, such as wearable biomedical devices and electronic health record, but a significant portion of these data are unlabelled. In order to provide the patient with real-time, accurate and reliable recommendations, the HDT needs to incorporate AI algorithms which can be efficiently trained by using these unlabelled data. In addition, instead of developing training-specific AI models for each task (e.g., tasks like glucose level prediction and providing healthy lifestyle recommendations), HDT should incorporates GAI for reducing redundant computations.

\subsection{Green HDT}
Implementation and evolution of HDT will require a lot of resources. To enhance users’ QoE, cloud-edge computing solutions will be essential to enhance interactions among various PTs and VTs, while providing real-time and precise health monitoring. Hence, the energy requirements to support massive communication and computation in HDT will continue to rise, exacerbating energy consumption while increasing environmental concerns such as greenhouse gas emissions. Additionally, as the report published by United Nations, HDT will pose serious risks for the sustainable development \cite{526}. To this end, green communication and computing services for HDT, termed as green HDT, are therefore imperative to achieve sustainability, although it is presently unclear how such an energy-efficient scheme can be achieved. New research directions needed to achieve green HDT include i) the development of new architectures with a focus on sustainability to support green cloud-edge networking and computing in HDT, ii) design and development of novel energy-efficient resource allocation schemes with the ability to support green HDT, and iii) integration of green HDT with other emerging green technologies for enhancing efficiency.

\emph{Use-case}: The maintenance of HDT on the network side requires significant resources (e.g., storage, communication and computing), which exacerbates the energy consumption and thereby leads to environmental concerns. Green HDT can be adopted to alleviate the energy burden resulted from the implementation of HDT on the network, through developing new energy-efficient architectures of HDT, designing green resource allocation schemes to increase the energy utilization efficiency, or integrating green HDT with other emerging green technologies, such as energy harvesting with renewable energy sources or energy storage and dispatch systems.

\subsection{Metaverse}
As a disruptive next-generation technology, Metaverse may revolutionize human lifestyles. 
Metaverse users are telepresent and immersed in a virtual world as human-like avatars, acting as VTs of their corresponding PTs. In addition to VTs of HDTs, the next-generation DT-enabled environment will also include other DTs, such as city DTs, unmanned aerial vehicle DTs, and satellite DTs. With an expected emergence of a massive diversity of DT objects in the Metaverse, malicious DTs and activities have become inevitable. For instance, VTs with a large volume of sensitive individual medical data may be at risk from malicious DTs, which can launch attacks to compromise VTs in the virtual world, causing significant destruction. To prevent this, HDT requires effective security and privacy mechanisms to protect VTs in such complex Metaverse environments. Recently, researchers have made the first attempt to use HDT technology to simulate the cognitive patterns of malicious hackers for proactive cybersecurity defense in the Metaverse \cite{521}.

\emph{Use-case}: Envision that in the Metaverse, a user with type 1 diabetes has a VT in the virtual world. Such VT monitors the user's blood sugar levels and insulin dosage in real-time, providing constant feedback to the user's healthcare professionals in the physical world. However, due to the lack of supervision, the VT may be compromised by a malicious DT in the Metaverse, which gains access to the user's medical data and begins tampering with the blood sugar readings. This potentially leads to fatal consequences for the user if it is not detected and corrected quickly. To prevent such attacks, the HDT in the Metaverse should be equipped with advanced security and privacy mechanisms. These mechanisms tailored to HDT in Metaverse detect anomalies, such as unexpected changes in the VT's behavior or data.

\section{Conclusion} \label{SE9}
In this survey, we have introduced the networking architecture of HDT in PH applications, and then, systematically surveyed the key supporting technologies for implementing such networking architecture.
We first present the motivation of HDT, along with its definition and core benefits. Then, we compare HDT with the conventional DTs, and further clarify that the traditional DT in healthcare is evolving toward HDT. After that, we illustrate its universal framework, essential functions, design requirements and challenges. After providing an overview of the networking architecture and key supporting technologies for HDT, we comprehensively review different data acquisition approaches (i.e., pervasive sensing and electronic health record), communication techniques (i.e., on-body and beyond-body communications), computing paradigms (i.e., multi-access edge computing and edge-cloud collaboration), data management processes (i.e., data pre-processing, data storage and data security and privacy), data analysis and decision making methods (i.e., artificial intelligence in various healthcare applications), all in the view of HDT and its end-to-end data stream processing procedure. Finally, we outline the future research directions of HDT.
\bibliographystyle{IEEEtran}
\bibliography{ref}

\newpage

\end{document}